\documentstyle[12pt,epsf,epsfig]{article}
\newcount\timecount
\newcount\hours \newcount\minutes  \newcount\temp \newcount\pmhours
\hours = \time
\divide\hours by 60
\temp = \hours
\multiply\temp by 60
\minutes = \time
\advance\minutes by -\temp
\def\hour{\the\hours}
\def\minute{\ifnum\minutes<10 0\the\minutes
            \else\the\minutes\fi}
\def\clock{
\ifnum\hours=0 12:\minute\ AM
\else\ifnum\hours<12 \hour:\minute\ AM
      \else\ifnum\hours=12 12:\minute\ PM
            \else\ifnum\hours>12
                 \pmhours=\hours
                 \advance\pmhours by -12
                 \the\pmhours:\minute\ PM
                 \fi
            \fi
      \fi
\fi
}

\def\monthname{\relax\ifcase\month 0/\or January\or February\or
   March\or April\or May\or June\or July\or August\or September\or
   October\or November\or December\else\number\month/\fi}

\def\bold#1{\setbox0=\hbox{$#1$}%
     \kern-.025em\copy0\kern-\wd0
     \kern.05em\copy0\kern-\wd0
     \kern-.025em\raise.0433em\box0 }

\def\beq{\begin{equation}}
\def\eeq{\end{equation}}

\def\ga{\mathrel{\raise.3ex\hbox{$>$\kern-.75em\lower1ex\hbox{$\sim$}}}}
\def\la{\mathrel{\raise.3ex\hbox{$<$\kern-.75em\lower1ex\hbox{$\sim$}}}}
\def\gev{{\rm \, Ge\kern-0.125em V}}
\def\tev{{\rm \, Te\kern-0.125em V}}
\def\gyr{{\rm \, G\kern-0.125em yr}}

\def\nl{\hfill\nonumber\\&&}
\def\nnl{\hfill\nonumber\\}

\def\gappeq{\mathrel{\rlap {\raise.5ex\hbox{$>$}}
{\lower.5ex\hbox{$\sim$}}}}
\def\lappeq{\mathrel{\rlap{\raise.5ex\hbox{$<$}}
{\lower.5ex\hbox{$\sim$}}}}
\def\Toprel#1\over#2{\mathrel{\mathop{#2}\limits^{#1}}}

\def\mchi{m_{\tilde \chi}}

\def\m12{m_{1\!/2}}

\def\PL{{\it Phys.Lett.} }
\def\PR{{\it Phys.Rev.} }

\begin{document}
\begin{titlepage}
\pagestyle{empty}
\baselineskip=21pt
\rightline{hep-ph/0308075}
\rightline{CERN--TH/2003-175}
\rightline{UMN--TH--2210/03}
\rightline{FTPI--MINN--03/21}
\vskip 0.2in
\begin{center}
{\large{\bf High-Energy Constraints on the Direct Detection \\
of MSSM Neutralinos}}
\end{center}
\begin{center}
\vskip 0.2in
{{\bf John Ellis}$^1$, {\bf Keith
A.~Olive}$^{2}$, {\bf Yudi Santoso}$^{2}$ and {\bf Vassilis C.~Spanos}$^{2}$}\\
\vskip 0.1in
{\it
$^1${TH Division, CERN, Geneva, Switzerland}\\
$^2${William I. Fine Theoretical Physics Institute,
University of Minnesota,  \\ Minneapolis, MN 55455, USA}}\\
\vskip 0.2in
{\bf Abstract}
\end{center}
\baselineskip=18pt \noindent

The requirement that the MSSM remain an acceptable effective field theory
up to energies beyond the weak scale constrains the sparticle spectrum,
and hence the permissible ranges of cold dark matter neutralino-proton
cross sections. Specifically, squarks are generally much heavier than
sleptons if no tachyons are to appear before the GUT scale $\sim
10^{16}$~GeV, or even before 10~TeV. We display explicitly the allowed
ranges of effective squark and slepton masses at the weak scale, and the
cross-section ranges allowed if the MSSM is to remain valid without
tachyons up to 10~TeV or the GUT scale. The allowed areas in the cross
section-mass plane for both spin-independent and spin-dependent scattering
are significantly smaller than would be allowed if the MSSM were required
to be valid only around the weak scale. In addition to a reduction in the
maximum cross section, the upper limit on the neutralino mass is greatly
reduced when tachyons are avoided, particularly for smaller values of the
squark masses.

\vfill
\leftline{CERN--TH/2003-175}
\leftline{August 2003}
\end{titlepage}
\baselineskip=18pt

\section{Introduction} 

At the present time, supersymmetry is a leading candidate for physics
beyond the Standard Model (SM). It provides a technical solution to the
problematic naturalness of the hierarchy of mass scales in particle
physics~\cite{hierarchy} and its renormalization-group equations (RGEs)
allow the SM parameters to be run up the Planck scale, enabling the gauge
couplings to unify at a high-energy GUT scale $\sim
10^{16}$~GeV~\cite{gut}. Furthermore, with $R$ parity conserved as in the
minimal supersymmetric extension of the SM (MSSM), the lightest neutralino
$\chi$, which is expected to be the lightest supersymmetric particle
(LSP), is a good candidate for providing the cold dark matter required by
astrophysics and cosmology~\cite{EHNOS}.

The MSSM contains two Higgs doublets and each SM particle has a
superpartner~\cite{mssm}. In order to obtain viable phenomenology,
supersymmetry should be broken by soft terms~\cite{softterms}. In the most
general case, there are more than 100 new parameters in this soft
supersymmetry-breaking sector, suggesting two approaches for doing the
MSSM phenomenology. The first approach is to assume that some simplified
pattern of supersymmetry breaking is input at the GUT scale, and then
evolve the RGEs for the MSSM down to lower energies, so as to calculate
the spectrum at the weak scale. The simplest model of this kind is the
CMSSM~\cite{cmssm}-\cite{cmssmmap}, in which the soft supersymmetry-breaking
masses $m_{1/2}$ of the gauginos and $m_0$ of the sfermions, as well as
the trilinear parameters $A_0$, are each assumed to be universal at the
GUT input scale. Extensive studies have been made of this and other GUT
input scenarios~\cite{nonu}-\cite{notthecmssm}.

Another approach is the low-energy effective supersymmetric theory (LEEST)  
approach. In this case, one does not care about any particular mass
relation at high energy scales, and simply uses the weak scale as the
input scale~\cite{lowmssm2}-\cite{lowmssm}. The only phenomenological
guidance used is provided by the constraints from experimental
observations. However, if one allows all the soft supersymmetry-breaking
parameters to be free, there will again be too many parameters for a
satisfactory phenomenological analysis to be possible, so this approach
usually also assumes some simplification at the weak scale, e.g.,
universal slepton and/or squark masses, in spite of the fact that these
universalities would not look natural from the GUT point of view.

One argument for using the LEEST approach is that experiments have not
been able to probe energy scales higher than 1 TeV, and no one really
knows what the higher-energy physics really is. However, loop corrections
should not be abandoned, unless we want to throw out Quantum Field Theory
altogether, and the RGEs are the best way to treat these. Even if one does
not impose any particular structure at some GUT scale, the RGE running
could in general lead to some sfermions becoming tachyonic at some higher
energy scale, i.e., $m_{\tilde f}^2(Q) < 0$ for some $Q > M_W$. To avoid
this tachyon problem, one must impose suitable positivity constraints on
the different sfermion masses in the LEEST.  These would depend on the
scale up to which the MSSM RGEs are thought to apply, above which one
might add some extra fields in a way suitable for avoiding the tachyonic
instability.  For example, in the CMSSM with non-universal Higgs masses (NUHM),
it was shown \cite{nuhm} that positivity constraints on the soft Higgs masses
at the GUT scale imposes strong constraints on the allowable parameter space.

The CMSSM and LEEST approaches differ in their treatments of this
tachyonic problem. If one uses the GUT scale as the input scale and makes
the CMSSM universality assumption, it is trivial to avoid the tachyon
problem by imposing relatively minor restrictions on the MSSM parameters $\mu, A_0$
and $\tan \beta$. In a LEEST analysis, the tachyon problem is accentuated
if one chooses spectra that optimize particular observables.  In
this paper, we discuss one particular example, that of the
neutralino-proton cross section, which is crucial for experiments
searching directly for cold dark matter particles via their scattering on
ordinary matter. In this case, one may obtain encouragingly large cross
sections by choosing relatively light squark masses. However, renormalization
effects are prone to driving such light squarks tachyonic at some energy
scale that is not far beyond the weak scale. We consider this contrary to
one of the primary ambitions of supersymmetry, which was to formulate a
plausible extension of the SM that could remain valid all the way up some
high energy scale, such as the GUT scale, and make the mass hierarchy
natural.

In view of the importance of this issue, we first summarize in Section~2
why we consider that tachyonic sfermions unacceptable. Then, in
Section~3, we exhibit some of the correlations among the observable
sparticle masses that arise if we use non-tachyonic soft
supersymmetry-breaking sfermion masses at the GUT or 10-TeV scale as
inputs to the MSSM. We emphasize the well-known point, for example, that the effective
squark masses at the weak scale must be much greater than the effective
slepton masses in this case, unless we start with slepton masses much
greater than squark masses at the input scale. Then, Section~4 outlines the
more general
model we use for our LEEST analysis, in which no hypothesis is made about
extrapolability to the GUT scale. Section~5 compares the elastic 
scattering cross sections 
found for different choices of the energy scale up to which the
LEEST is postulated to apply without tachyons appearing. Finally,
Section~6 presents our conclusions.

\section{Vacuum Instability in the MSSM}

Let us first consider a sample RGE for the case of a right-handed 
up-type squark:
\beq
Q {d m^2 \over dQ} \simeq {1 \over 8 \pi^2} \left[ -{16 \over 3} g_3^2 M_3^2 -
{16 \over 9} g_1^2 M_1^2 \right]
\label{RGE}
\eeq
where $g_{1(3)}$ is the $U(1)_Y(SU(3)_c)$ gauge coupling
and $M_{1(3)}$ the corresponding gaugino mass. 
At the one-loop level, (\ref{RGE}) can be solved exactly using the 
solutions
to the gauge coupling RGEs and the relationship between ratios of
gaugino masses and gauge couplings, and yields~\cite{FORS}
\begin{eqnarray}
m^2(Q) =  m^2(M_W) &-& \frac{2}{3 \pi^2} g_3^2(M_3) M_3^2 \ln (Q/M_3) \left\{
\frac{ 1+ 3 g_3^2(M_3) \ln(Q/M_3)/(16\pi^2)}{[ 1+3g_3^2(M_3) \ln
(Q/M_3)/(8\pi^2)]^2} \right\} \nnl
&-&  \frac{2}{9 \pi^2} g_1^2(M_1) M_1^2 \ln (Q/M_1) \left\{
\frac{ 1 -11 g_1^2(M_1) \ln(Q/M_1)/(16\pi^2)}{[ 1 -11 g_1^2(M_1) \ln
(Q/M_1)/(8\pi^2)]^2} \right\}
\end{eqnarray}
{}From this equation, it is straightforward to see the monotonic decrease
in the squark mass at $Q > M_W$. In addition, one readily sees that as the
gaugino mass (and in particular the gluino mass) {\it increases}, the rate
at which the squark mass decreases {\it also increases}.  Therefore,
depending on one's choice of $m^2(M_W)$, it is quite generic to find that
$m^2(Q_0) = 0$, for some $M_W < Q_0 < M_{GUT}$.

Problems with tachyonic
instabilities were first raised in this context in~\cite{FORS}.
The resulting tachyonic behaviour of the solution to the RGEs is an 
indication that 
the scalars are being expanded about the wrong vacuum. 
In this case, there are at least two options for treating the problem. 
\begin{enumerate}
\item One can introduce new physics at an energy scale below $Q_0$, 
so that the RGE running is affected and the positivity of $m^2$ is maintained.
Of course, such a solution is by fiat outside the realm of the MSSM and 
hence of this paper. 
\item If no new physics is introduced below $Q_0$, then it must be 
present at some higher energy $Q > Q_0$ - perhaps the GUT scale or some 
intermediate scale - in order to restore the stability of the theory.  In this 
case, however, one expects the squarks to acquire non-zero vacuum expectation 
values $v(Q) \simeq Q$, related to the new very high-energy scale $Q > Q_0$.
The severity of such instabilities and the existence of new vacua in the
presence of flat directions in the effective potential was particularly
stressed in~\cite{FORS}, and explicit solutions with $v(Q) \simeq Q$ were
exhibited. 
\end{enumerate}

To see that the broken vacuum persists down to low energies, consider the
evolution of the RGEs at large field values.  Any field coupled to the
squark with the large vacuum expectation value becomes massive, and hence
is unavailable for the radiative running of $m^2$ back to positive values.
In this case, the solution to the RGEs will have $m^2(Q) < 0$ for all $Q$
and the theory will possess a low-energy global minimum with charge and
color breaking (CCB)~\footnote{For discussions concerning CCB vacua in the
CMSSM, see~\cite{ccb}.}.
 
It was
argued in~\cite{RR,KLS} that this situation could be tolerated, because the
Universe would naturally fall into our false vacuum as the cosmological
temperature fell, and then take much longer than the present age of the
Universe to tunnel into the true vacuum. We find this point of view
disturbingly anthropic. Moreover, it has been observed that, if the early
Universe went through an inflationary phase, it is likely to have fallen
directly into the true vacuum, so the lifetime of the false vacuum is
irrelevant~\cite{FORSS}. 

Following this debate, cosmological data ($\Omega_{tot} \simeq 1$,
near-scale-invariant adiabatic density fluctuations that appear Gaussian,
etc.~\cite{WMAP}.) have come increasingly to favour some variant of
inflationary cosmology. At best, the stability conditions on the MSSM
might depend on the details of the inflationary cosmology chosen. At
worst, their neglect lead to an entirely inconsistent theory.  Overall, we
think that tolerating tachyonic sfermions would court disaster.
Furthermore, since the tachyonic Higgs instability induced by the top
Yukawa coupling, leading to dynamical electroweak symmetry breaking, is
such an attractive feature of the MSSM, we are reluctant to introduce
ambiguity into the mechanism by accepting other tachyonic instabilities in
the MSSM.

\section{GUT Correlations Among MSSM Parameters}

The simplest way to avoid the tachyonic instabilities discussed above is
to input acceptable parameters at the high-energy scale, as is done in the
CMSSM. Assuming input soft supersymmetry-breaking parameters at the GUT or
some other high-energy scale, one may use the MSSM RGEs to calculate the
sparticle mass spectrum at the weak scale. The RGE evolution leads in
general to correlations among the various supersymmetric parameters.
Moreover, in some cases, a wide range of these input parameters at the GUT
scale corresponds to a specific range down at the weak scale, as we now
demonstrate.

We assume in our analysis the following ranges of the sparticle mass 
parameters at the GUT scale:
\begin{eqnarray}
&& {m_{\tilde{Q}}}_0 = {m_{\tilde{U}}}_0= {m_{\tilde{D}}}_0   \leq 2 \tev \, ,
\quad 
   {m_{\tilde{L}}}_0 = {m_{\tilde{E}}}_0   \leq 2 \tev , \nonumber \\
&& m_{1/2} \leq 2 \tev  \, , \quad  
   m_{1,2} \leq 2 \tev , \nonumber \\
&& |A_0| \leq 1 \tev \, , \quad \tan\beta  \leq 58,
\label{eq:range_cor}
\end{eqnarray}
where ${m_{\tilde{Q},\tilde{U},\tilde{D}}}_0$ and
${m_{\tilde{E},\tilde{L}}}_0$ are the input GUT-scale soft
supersymmetry-breaking masses for all three families of squarks and
sleptons, respectively, $m_{1/2}$ is the common gaugino mass, and
$m_{1,2}$ are the soft supersymmetry-breaking masses for the two Higgs
doublets.

In the CMSSM, one postulates GUT unification relations among the soft
supersymmetry-breaking masses of the sfermions and Higgs bosons:
${m_{\tilde{Q}}}_0 = {m_{\tilde{U}}}_0 = {m_{\tilde{D}}}_0
={m_{\tilde{L}}}_0 = {m_{\tilde{E}}}_0 = m_{1,2}$.  These conditions are
relaxed here.  Even so, as we now show, the RGE evolution of the soft
supersymmetry-breaking parameters between the GUT and weak scales result
in some striking correlations between the weak-scale parameters.
Therefore, one cannot treat their values at the weak scale as completely
independent quantities, if the model under study corresponds to a more
fundamental theory at a higher scale.  To illustrate these correlations,
we have chosen to present a few particular examples. We assume
universality between the SU(3), SU(2) and U(1) gaugino masses $m_{1/2}$ at
the GUT scale and discuss the dependences of the weak-scale values of
various soft parameters in the sfermion sector. Specifically, we display
the dependences of the soft supersymmetry-breaking sfermion masses and
trilinear scalar couplings on the common gaugino mass $m_{1/2}$ at the
input GUT scale.

\begin{figure}
\begin{center}
\mbox{\epsfig{file=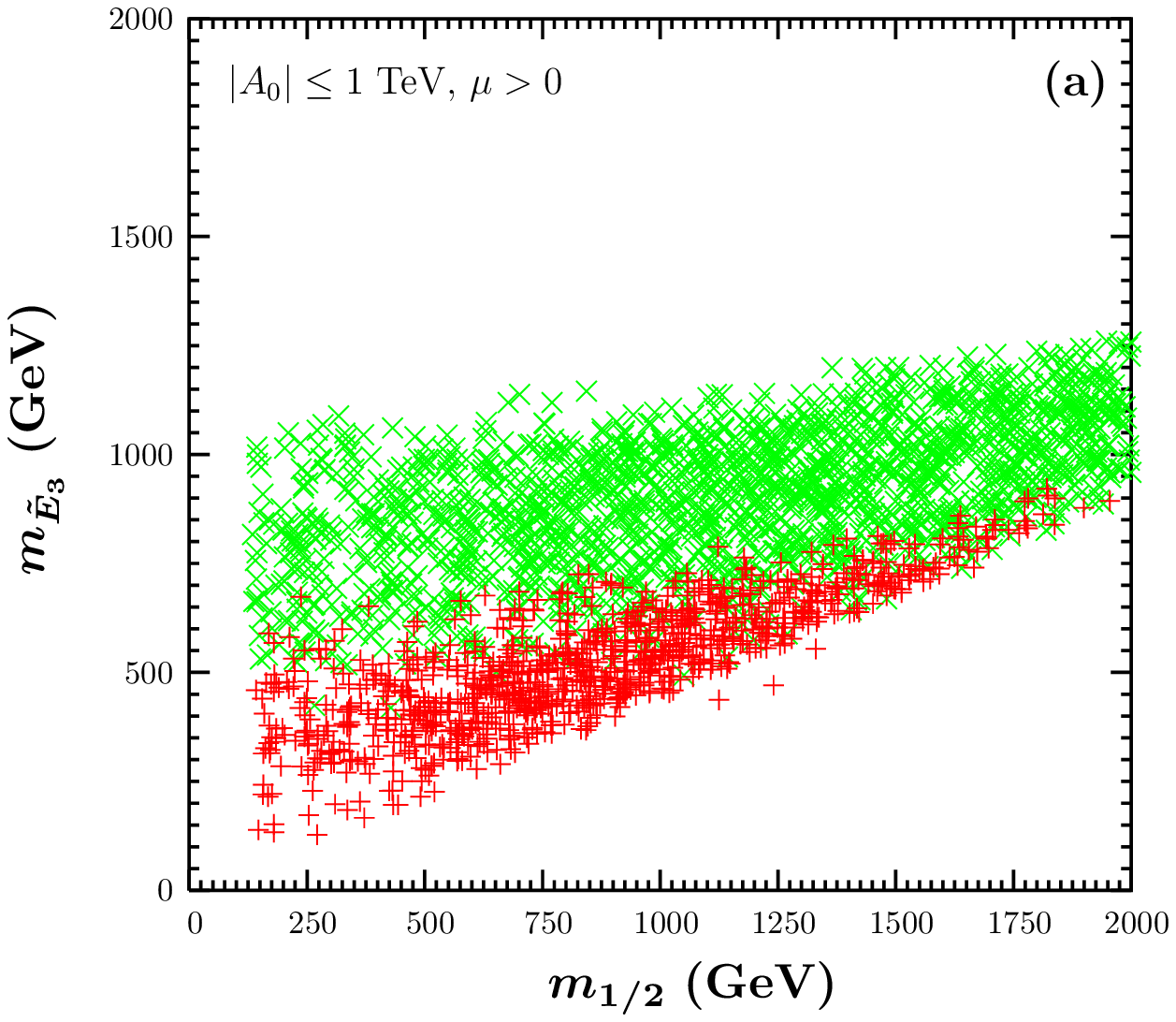,height=6.5cm}}
\hspace{0.2in}
\mbox{\epsfig{file=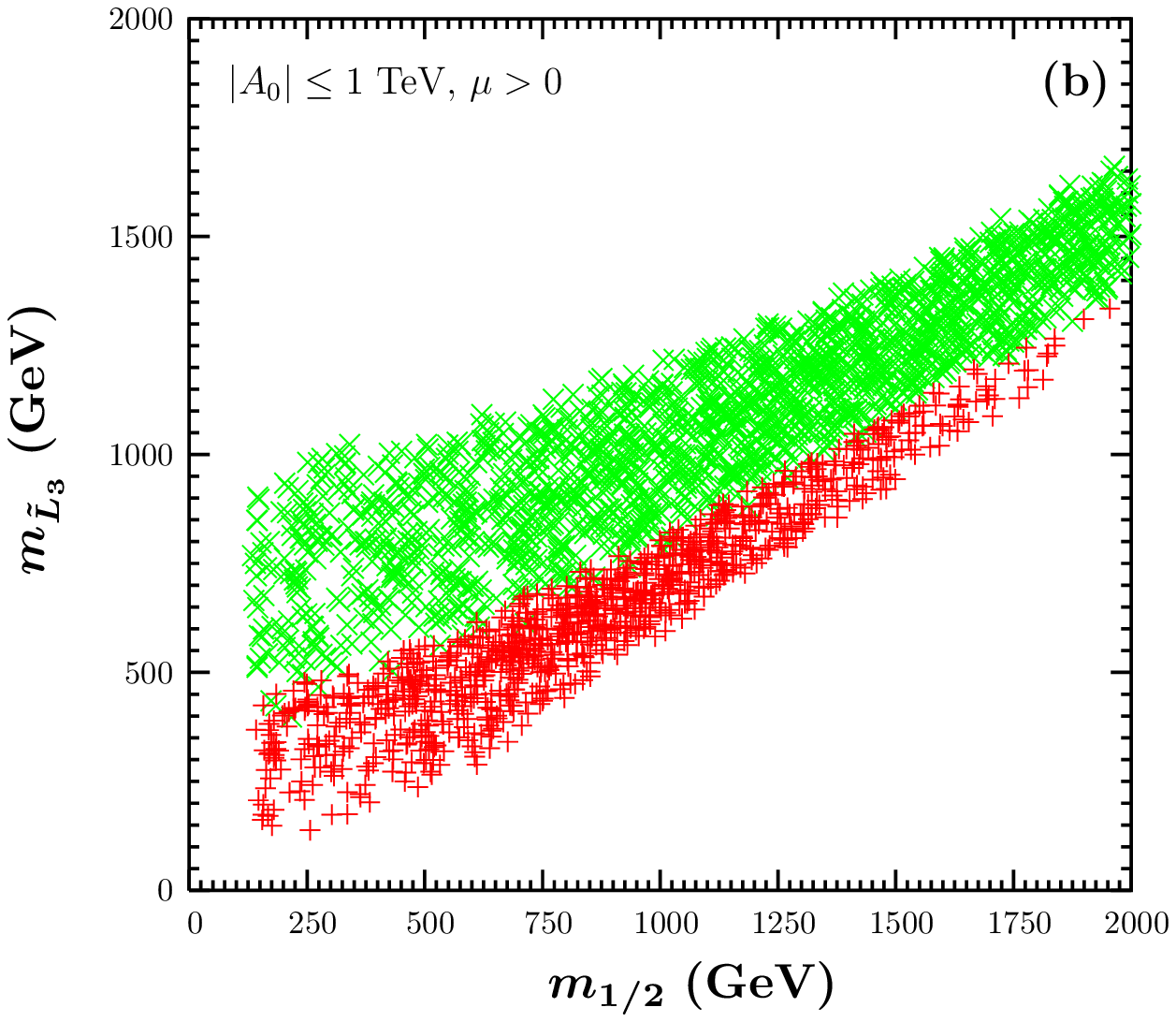,height=6.5cm}}
\vskip 0.4in
\mbox{\epsfig{file=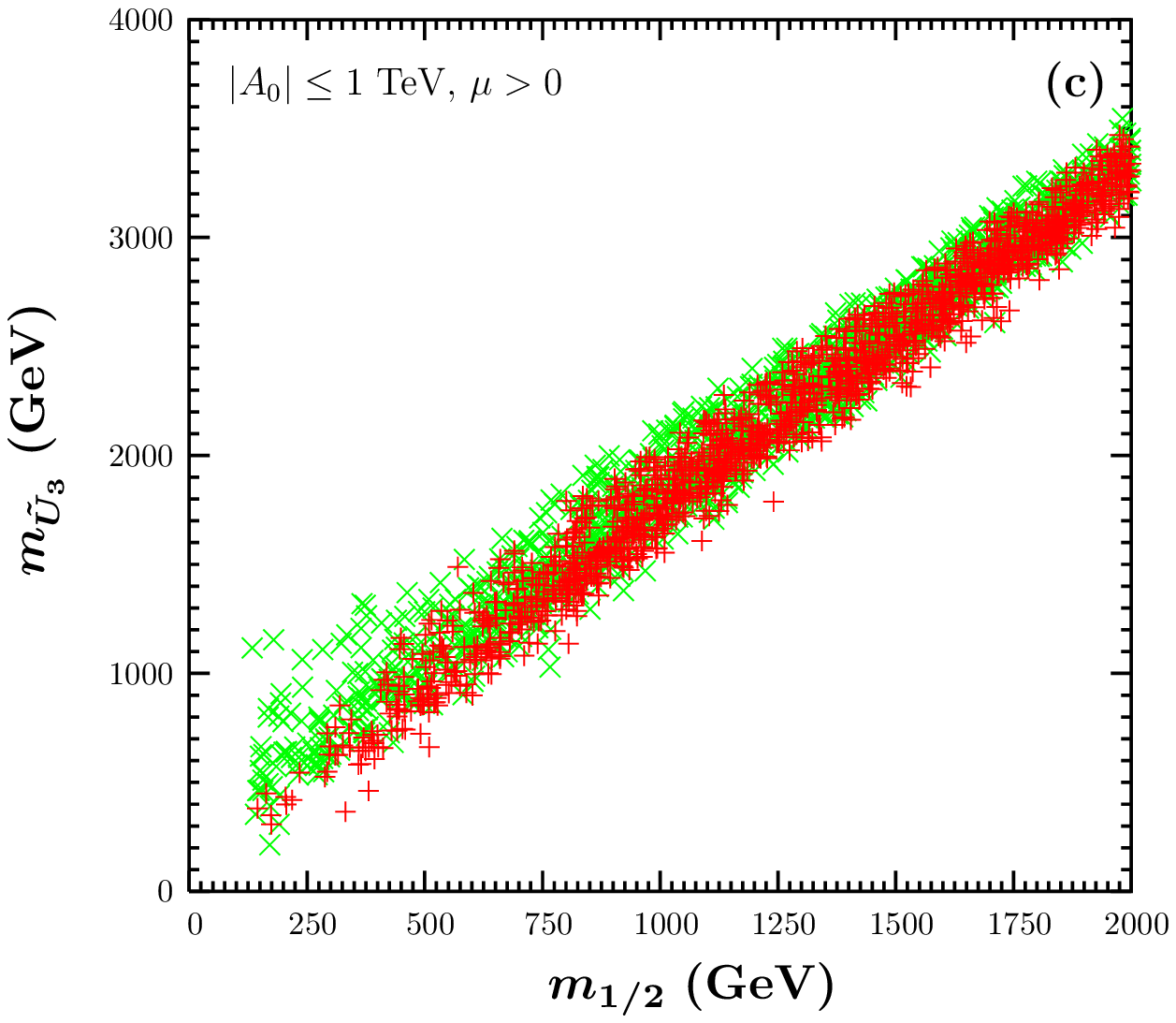,height=6.5cm}}
\hspace{0.2in}
\mbox{\epsfig{file=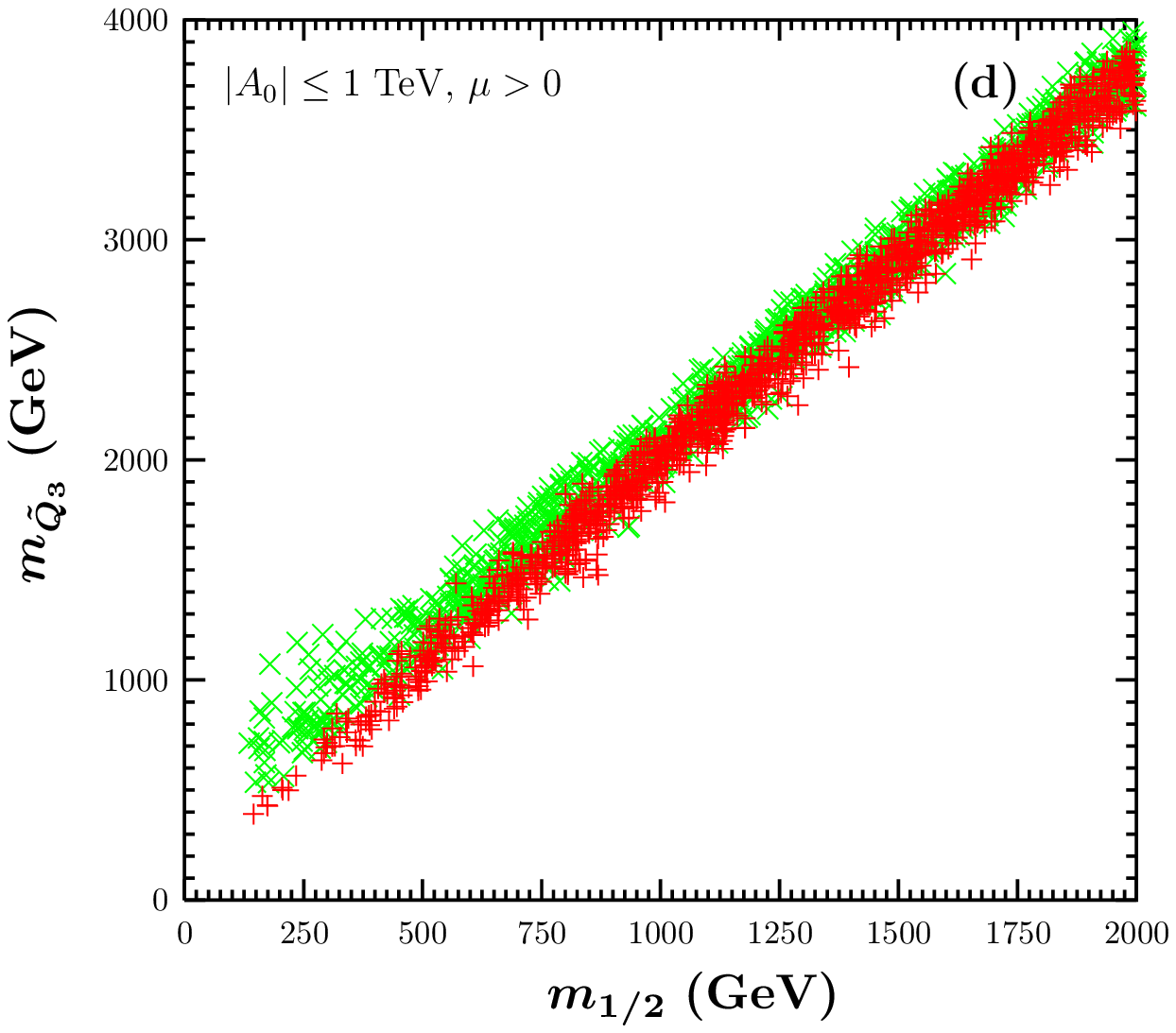,height=6.5cm}}
\end{center}
\vskip 0.2in
\caption{\label{fig:msql}\it
The weak-scale values of the soft supersymmetry-breaking mass parameters 
(a) $m_{\tilde{E}_3}$, (b) $m_{\tilde{L}_3}$, (c) $m_{\tilde{U}_3}$ and 
(d) $m_{\tilde{Q}_3}$, as functions of the 
universal gaugino mass $m_{1/2}$ input at the GUT scale.
We use light (green) $\times$ signs / dark (red) $+$
signs to denote points with ${m_{\tilde{E}}}_0$ or ${m_{\tilde{Q}}}_0 \leq 500 
\gev$ / between $500 \gev$ and $1000 \gev$ input at the GUT scale. 
}
\end{figure}

We first show in Figure~\ref{fig:msql} the weak-scale values of the soft
supersymmetry-breaking mass parameters (a) $m_{\tilde{E}_3}$ and (b)
$m_{\tilde{L}_3}$, which appear in the slepton sector of the third
generation, and (c) $m_{\tilde{U}_3}$ and (d)  $m_{\tilde{Q}_3}$, which
appear in the squark sector of the third generation, as functions of
$m_{1/2}$. We use a random 50,000-point sample in the range of parameters
(\ref{eq:range_cor}), with $|A_0|<1$ TeV. The different symbols: [light
(green) $\times$ signs/ dark (red) $+$ signs] correspond to imposing the
constraints ${m_{\tilde{E}}}_0 \leq 500 \gev$ / between $500$ and $1000
\gev$ for Figs.~\ref{fig:msql}(a, b), and ${m_{\tilde{Q}}}_0 \leq 500
\gev$ / between $500$ and $1000 \gev$ in Figs.~\ref{fig:msql}(c, d),
respectively.

It is immediately apparent that the soft supersymmetry-breaking masses
are correlated, especially for large values of the gaugino mass, where
the corresponding terms dominate the RGEs for $m_{\tilde{E}_3}$ and
$m_{\tilde{L}_3}$. 
This can be seen from the approximate one-loop solutions to the RGEs
for the sleptons (we neglect the D-term contribution for clarity):
\begin{eqnarray}
m_{\tilde{E}_3}^2 & = & {m_{\tilde{E}_3}}_0^2 + 0.15 m_{1/2}^2, 
\nonumber \\
m_{\tilde{L}_3}^2 & = & {m_{\tilde{L}_3}}_0^2 + 0.52 m_{1/2}^2.
\label{lr1}
\end{eqnarray}
The correlation between the soft
supersymmetry-breaking masses $m_{\tilde{Q}_3}$, $m_{\tilde{U}_3}$ and
the gaugino mass, presented in Figure~\ref{fig:msql}(c, d), is even
tighter, due to the larger value of the combination $g_3M_3$ relative
to either $g_1M_1$ or $g_2M_2$, which drives the evolution of the slepton 
masses and is shown in the following approximate solution:
\begin{eqnarray}
m_{\tilde{U}_3}^2 & = & 0.52 {m_{\tilde{U}_3}}_0^2 + 5.4 m_{1/2}^2 + 
m_t^2, \nonumber \\
m_{\tilde{Q}_3}^2 & = & 0.04 {m_{\tilde{Q}_3}}_0^2 + 4.2 m_{1/2}^2 + 
m_t^2.
\label{lr2}
\end{eqnarray}
In this case, the substantial size of the top Yukawa coupling, which
appears in the RGE of $m_{\tilde{Q}_3}$, also plays an important role.  
The boundaries of the allowed regions in Fig.~\ref{fig:msql} can easily be
understood from (\ref{lr1}) and (\ref{lr2}).  The upper edge is clearly an
artifact of the choice of the upper cut-off on the GUT-scale masses. If
one were to select higher ranges for the soft supersymmetry-breaking
masses, i.e., greater than 1000 GeV for the input masses, the correlation
region would just move upwards to larger low-energy masses.  What is
important to note, however, is that the lower right parts of these
figures, regions with low mass sfermions and large gaugino masses, are
never populated.  The lower edge is, in fact, determined by the gaugino
mass contribution to the weak-scale masses. Indeed, we see that, because
the contributions to the weak-scale masses from the input squark masses
are much weaker than those from the input slepton masses (0.52 and 0.04
versus 1.0) the correlation is much tighter for squarks than for sleptons,
particularly for the left-handed stops. Furthermore, one sees that the
slope of the correlation is fixed by the coefficients of $m_{1/2}$, which
are much larger for squarks than for sleptons.

We note that, for any given value of $m_{1/2}$, the allowed ranges of
$m_{\tilde{L}_3}$ and $m_{\tilde{E}_3}$ at the weak scale are lower than those of
$m_{\tilde{Q}_3}$ and $m_{\tilde{U}_3}$, reflecting the greater
renormalizations of the squark mass parameters, as seen by comparing the 
coefficients of 
$m_{1/2}$ in (\ref{lr1}) and (\ref{lr2}). Several aspects of the
MSSM spectrum enter in the calculation of the relic LSP density, including
the $Z$ and Higgs masses as well as sfermion masses. However, in most of
the region where an acceptable relic mass density is obtainable in the
CMSSM, $\chi -$ slepton coannihilations are very important \cite{coann,moreco}. These are
controlled, in particular, by $m_{\tilde{E}_3}$ and $m_{\tilde{L}_3}$,
whereas the $\chi - p$ elastic scattering cross section is largely
controlled by squark masses. We see in panels (a) and (b) of
Figure~\ref{fig:msql} that these slepton mass parameters are strongly
correlated. Moreover, they are much smaller than the squark masses shown
in panels (c) and (d), implying that the $\chi - p$ elastic scattering
rate must be much smaller than would have been the case if the slepton and
squark masses were comparable.

As noted above, the squark and (to a lesser extent) the slepton masses
cannot be very small, unless $m_{1/2}$ is small. Within the GUT approach,
for example, if $m_{1/2} \simeq 1000$~GeV, then $m_{\tilde{Q}_3} \gappeq
2000$~GeV, with an uncertainty $\sim 10$~\%. Assuming small squark masses
but large $m_{1/2}$, as might be done within the LEEST approach to 
yield relatively large cross sections for large LSP masses, would lead to
the squarks becoming tachyonic long before the GUT scale.
For another discussion of `perverse' models with tachyonic squarks at the 
GUT scale, see \cite{Kane:2003iq}.

\begin{figure}
\begin{center}
\mbox{\epsfig{file=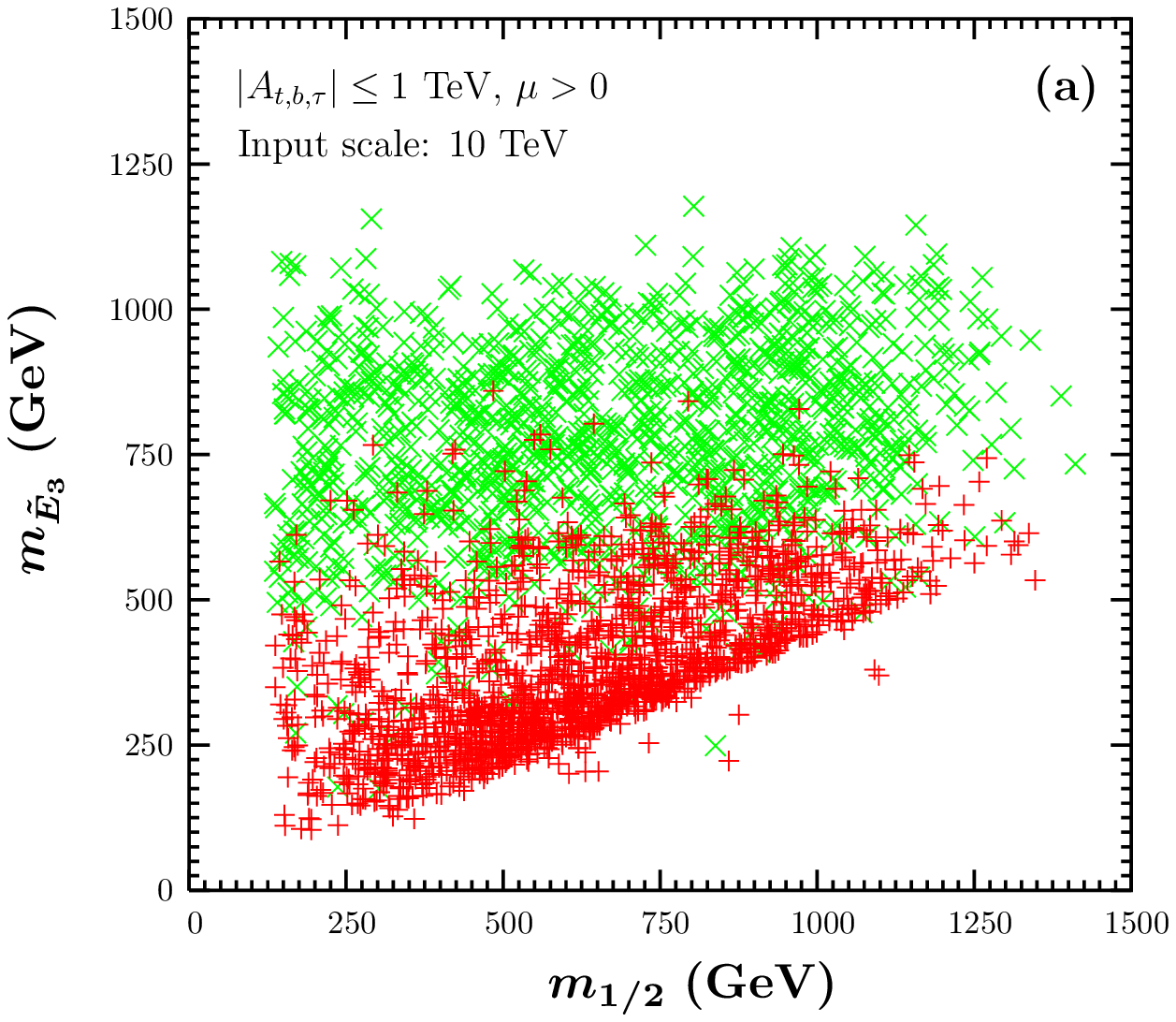,height=6.5cm}}
\hspace{0.2in}
\mbox{\epsfig{file=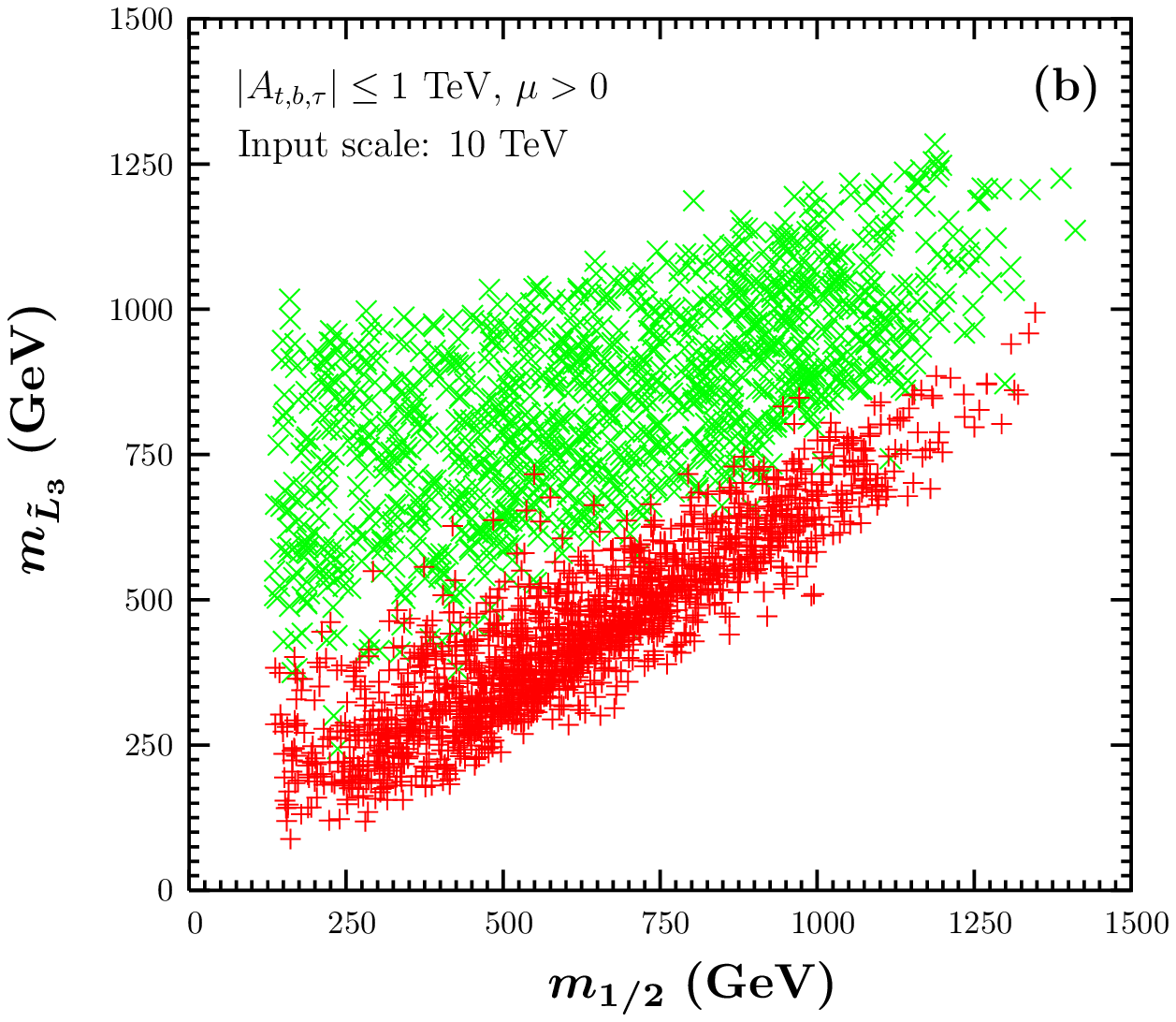,height=6.5cm}}
\vskip 0.4in
\mbox{\epsfig{file=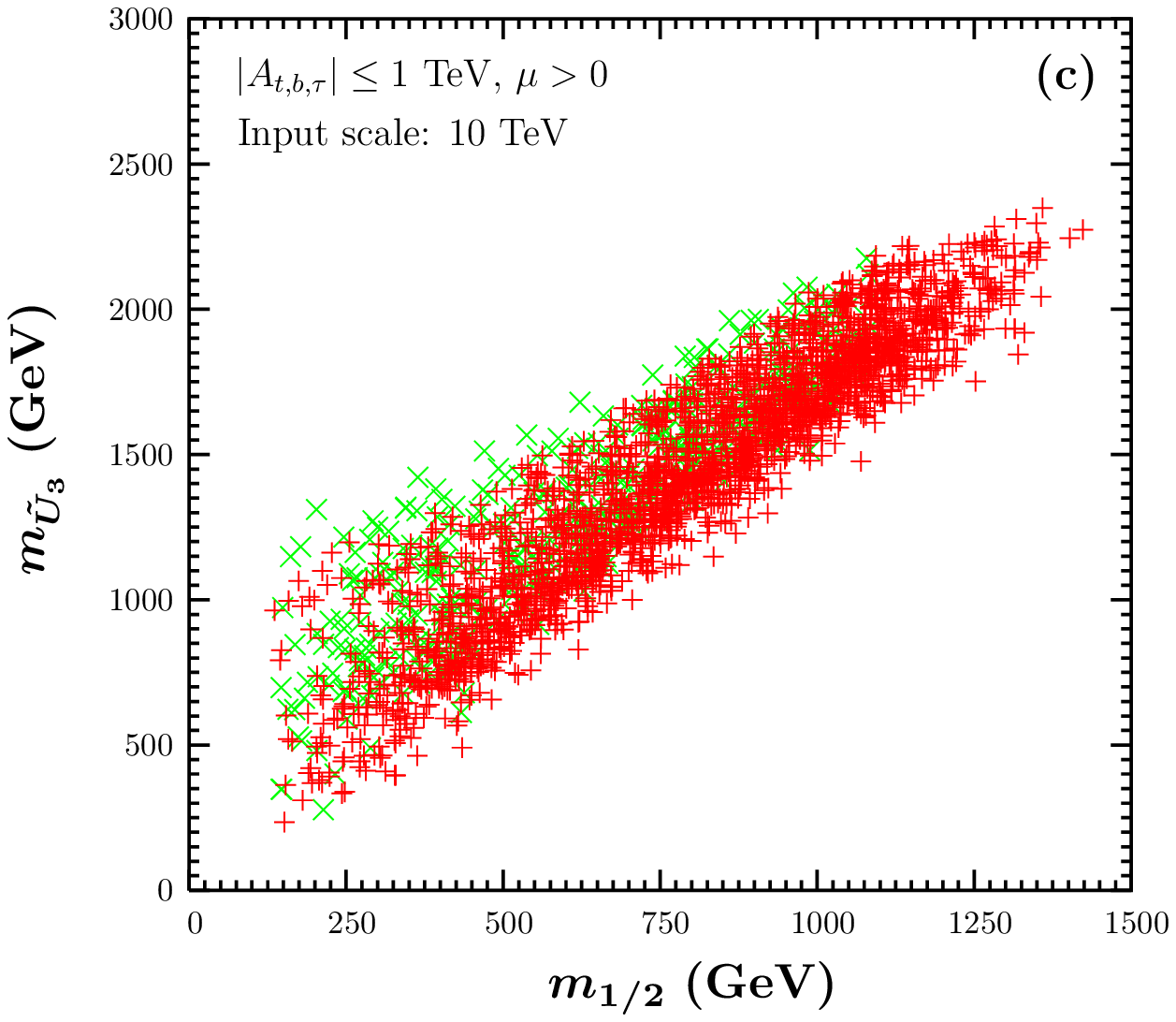,height=6.5cm}}
\hspace{0.2in}
\mbox{\epsfig{file=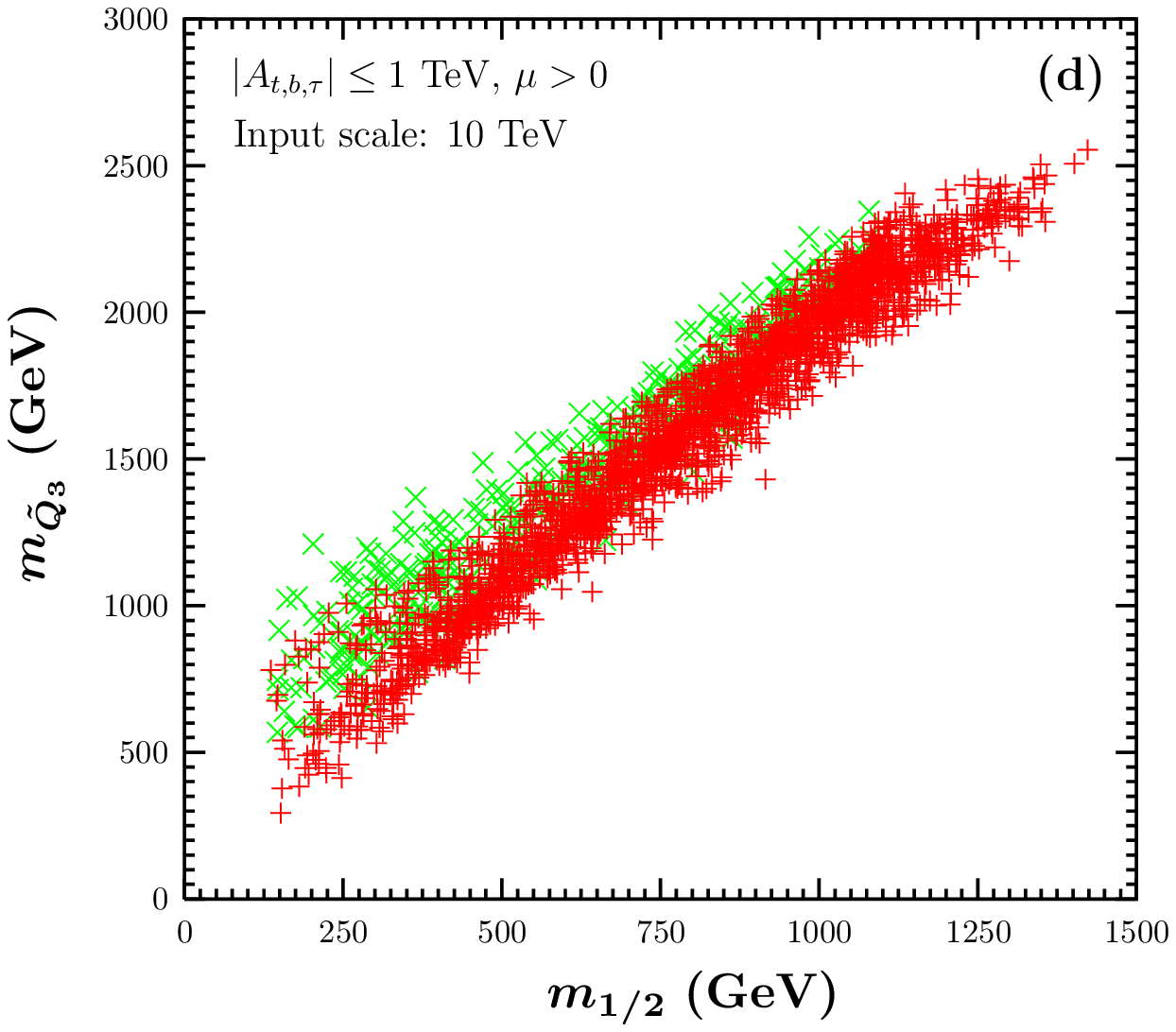,height=6.5cm}}
\end{center}
\vskip 0.2in
\caption{\label{fig:msql10}\it
The same quantities as in Fig.~\protect\ref{fig:msql} are plotted
as functions of the
universal gaugino mass $m_{1/2}$ input at the GUT scale. However, in these 
plots we use values of the soft supersymmetry-breaking masses and 
trilinear couplings input at 10~TeV.
}
\end{figure}

In Fig.~\ref{fig:msql10}, we exhibit similar correlations between the
weak-scale soft supersymmetry-breaking masses in the case where the same
ranges for the sfermion masses and trilinear couplings as in
(\ref{eq:range_cor}) are now assumed at a scale of 10~TeV. However, we
still assume GUT unification for the gaugino masses. In this case, we 
still observe correlations between $m_{1/2}$ and the weak-scale sfermion 
masses and trilinear couplings, though they are somewhat weaker than in 
Fig.~\ref{fig:msql}, where we had assumed (\ref{eq:range_cor}) at the GUT 
scale. As before, we also observe that the weak-scale squark masses are 
generally significantly larger than the corresponding slepton masses. 
For example, if $m_{1/2} \sim 1000$~GeV, the range of possible slepton 
masses still has no overlap with the range of possible squark masses. As 
already emphasized, if the physical squark and slepton masses are small and
(nearly) equal, then, except possibly for small values of $m_{1/2}$, the
squarks would become tachyonic at some scale not far above 
the weak scale~\footnote{This statement depends on our assumption of
universal gaugino
masses at the GUT scale. If one assumed non-universal gaugino masses 
$M_{1,2,3}$, one would
need $M_3 \ll M_{1,2}$ to get nearly equal squark and slepton physical
masses. This situation might not be suitable for providing dark 
matter.}.

Using these representative examples as a guide, we
conclude, that even if one abandons the unification of the soft
supersymmetry-breaking sfermion and Higgs-boson masses at the GUT scale,
RGE evolution yields interesting correlations among the various parameters
of the model also down at the weak scale.  The fact that certain
combinations of sfermion masses at the weak scale and gaugino masses 
are never realized in a top-down analysis reflects the 
tachyonic constraint that we will apply in the following section.

\section{Low-Energy Effective Supersymmetric Theory}

For our low-energy effective supersymmetric theory (LEEST) analysis, 
we use as our input parameters
the SU(2) gaugino mass $M_2$, the Higgs mixing parameter $\mu$, the
(assumed to be) universal soft supersymmetry-breaking slepton mass
$m_{\tilde \ell}$, the universal squark mass $m_{\tilde q}$, the trilinear
couplings $A_t$, $A_b$ and $A_\tau$, the pseudoscalar Higgs-boson mass
$m_A$, and $\tan \beta$.  We consider the following ranges for these
parameters at the weak scale:
\begin{eqnarray}
&& 100 \, \gev \leq M_2 \leq 2 \, \tev , \nl
100 \, \gev \leq m_{\tilde \ell} \leq 2 \, \tev , \nl
200 \, \gev \leq m_{\tilde q} \leq 4 \, \tev , \nl
90 \, \gev \leq m_A \leq 2 \, \tev , \nl
100 \, \gev \leq |\mu| \leq 2 \, \tev . 
\label{reflowmssm}
\end{eqnarray}
We analyze in this paper only the $\mu > 0$ case, fix $A_t = A_b = 1 \,
\tev$, assume $A_\tau = 0$, and sample $\tan \beta = 10$, 35 and 50. These
assumptions have been chosen to resemble those of~\cite{lowmssm}. We note 
that other LEEST analyses~\cite{lowmssm2} are more flexible,
allowing $m_{\tilde \ell} \simeq m_{\tilde q}$ with both small (i.e. $<
200$~GeV), a 
possibility not included in (\ref{reflowmssm}). However, with regard to the
discussion in the 
previous two sections, inclusion of the lower squark masses will almost
certainly lead to tachyonic instabilities making the choice (\ref{reflowmssm}) 
somewhat better motivated.  As usual, 
to obtain the physical sfermion masses, we add $D$-term corrections and
diagonalize the mass matrices.  The other gaugino masses, $M_1$ and $M_3$,
are approximated by $M_1 = 5/3 M_2 \tan^2 \theta_W$ and $M_3 = 29.74 \, M_2 \, 
\alpha_s$, so as to mimic gaugino unification at the GUT scale.

We should emphasize that this is an oversimplified model. First of all,
there is no known fundamental principle why all the slepton and all the
squark soft supersymmetry-breaking masses should each be universal at the
weak scale~\footnote{On the other hand, the experimental upper limits on
flavour-changing neutral interactions suggest that the physical masses of
particles with the same internal quantum numbers in different generations
should be relatively degenerate~\cite{EN}.}. Moreover, physical masses
should be calculated at their appropriate scale, and sfermion masses at
$M_Z$ could be significantly different from those at $M_{susy}$.
Nevertheless, this toy model is often used in the
literature~\cite{lowmssm2,lowmssm3,lowmssm}, and may be sufficient to
represent a general analysis of the effective MSSM.  It should be noted
that, in this kind of analysis, one must make some simplifying
assumptions, because otherwise the number of parameters is impractically
large.

We apply the constraints on new particles from direct LEP searches, namely
$m_{\chi^\pm} > 104$~GeV~\cite{LEPsusy}, $m_{\tilde e} >
99$~GeV~\cite{LEPSUSYWG_0101} and Higgs mass limits~\cite{LEPHiggs}. We
require the branching ratio for $b \rightarrow s \gamma$ to be consistent
with the experimental measurements~\cite{bsg}. We do not impose the
constraint suggested by $g_\mu - 2$ \cite{newBNL}, as the magnitude of the
Standard Model contribution is still undetermined, in view of the
discrepancy between the $e^+ e^-$ data and the $\tau$-decay
data~\cite{Davier}. For the relic density of neutralinos $\chi$, we
require that $0.1 \le \Omega_\chi h^2 \le 0.3$.  We use this more
conservative range rather than that suggested by the recent WMAP
data~\cite{WMAP} ($0.094 \le \Omega_\chi h^2 \le 0.129$), because it is
more suitable for comparison with previous work~\footnote{In fact, we have
found that restricting MSSM so that the relic density falls within the
WMAP range does not reduce significantly the ranges of the cross sections
possible for any given value of $m_\chi$, though it may restrict the range
of $m_\chi$ itself.}.  For studies on the effect of the new WMAP density in 
CMSSM models,  see \cite{cmssmmap}.

We emphasize the importance of implementing correctly the available
experimental constraints. For example, the SM limit of
114~GeV~\cite{LEPHiggs} applies in the CMSSM, and also in the LEEST
framework discussed here when $m_A$ is large. For small values of $m_A$
for which the $ZZh$ coupling factor $\sin (\beta - \alpha)$ is less than a
half, we relax the Higgs mass bound to 87 GeV. One might obtain much
larger elastic scattering cross sections if one used a smaller lower limit
on $m_h$. Similarly, the $b \to s \gamma$ constraint must be implemented
taking careful account of the theoretical as well as the experimental
errors, since it provides important lower limits on sparticle masses, in
particular for $\mu < 0$ and high values of $\tan \beta$.

\section{Elastic Neutralino-Proton Scattering Cross Sections}

Finally, we are ready to compute the elastic $\chi-p$ scattering cross 
section in the LEEST
framework, using a scan in 100-GeV steps of the parameter space 
over the ranges given in (\ref{reflowmssm}). 
For the points that pass all the phenomenological constraints mentioned 
earlier, we calculate the
neutralino-proton cross section $\sigma_{\chi p}$ using the procedure
described in~\cite{npcross}. We begin with the spin-independent (scalar)
part of the cross section, and then turn to the spin-dependent part.

\begin{figure}
\begin{center}
\vskip -0.5in
\mbox{\epsfig{file=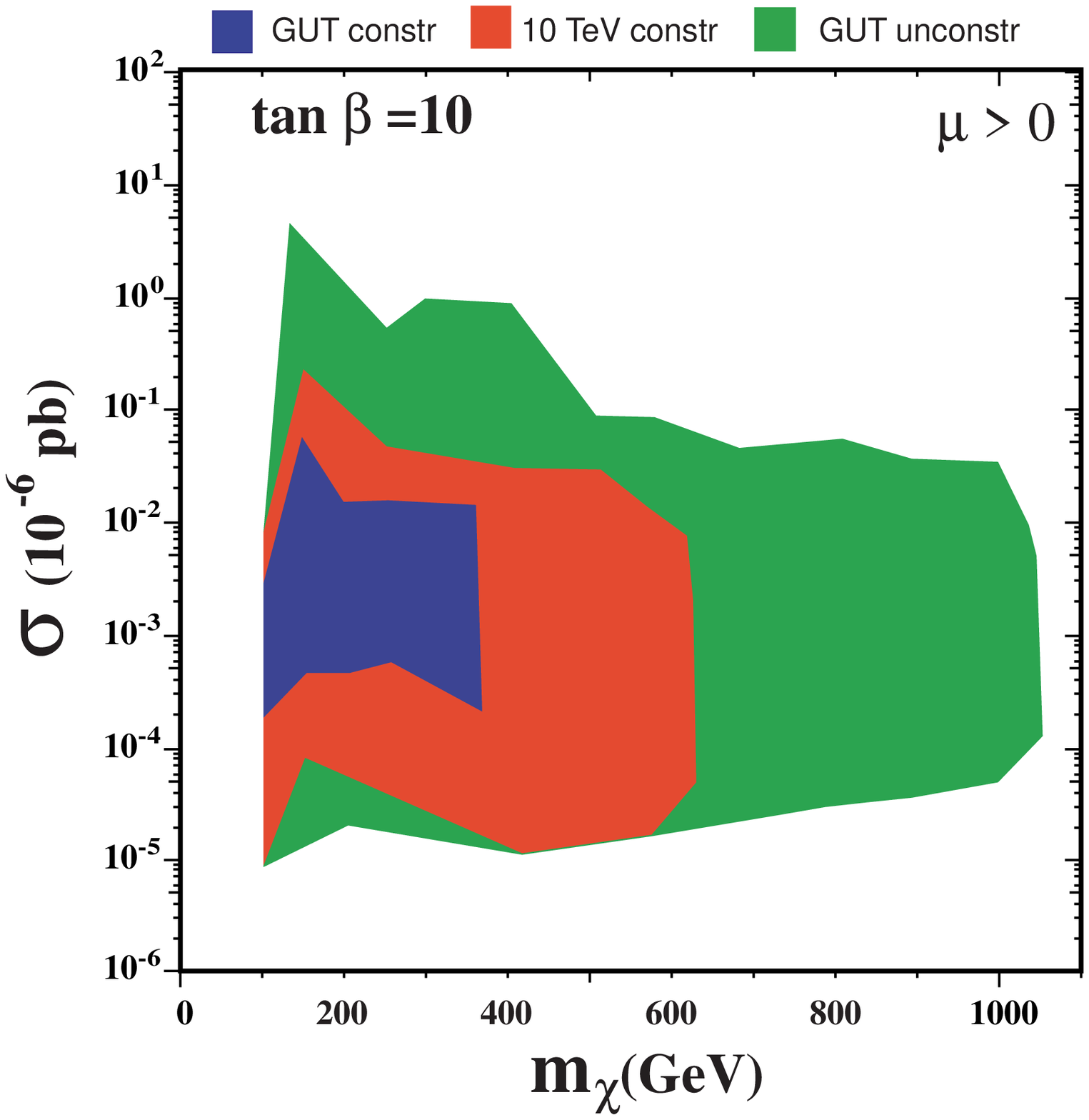,height=6.4cm}}
\hspace{0.4in}
\mbox{\epsfig{file=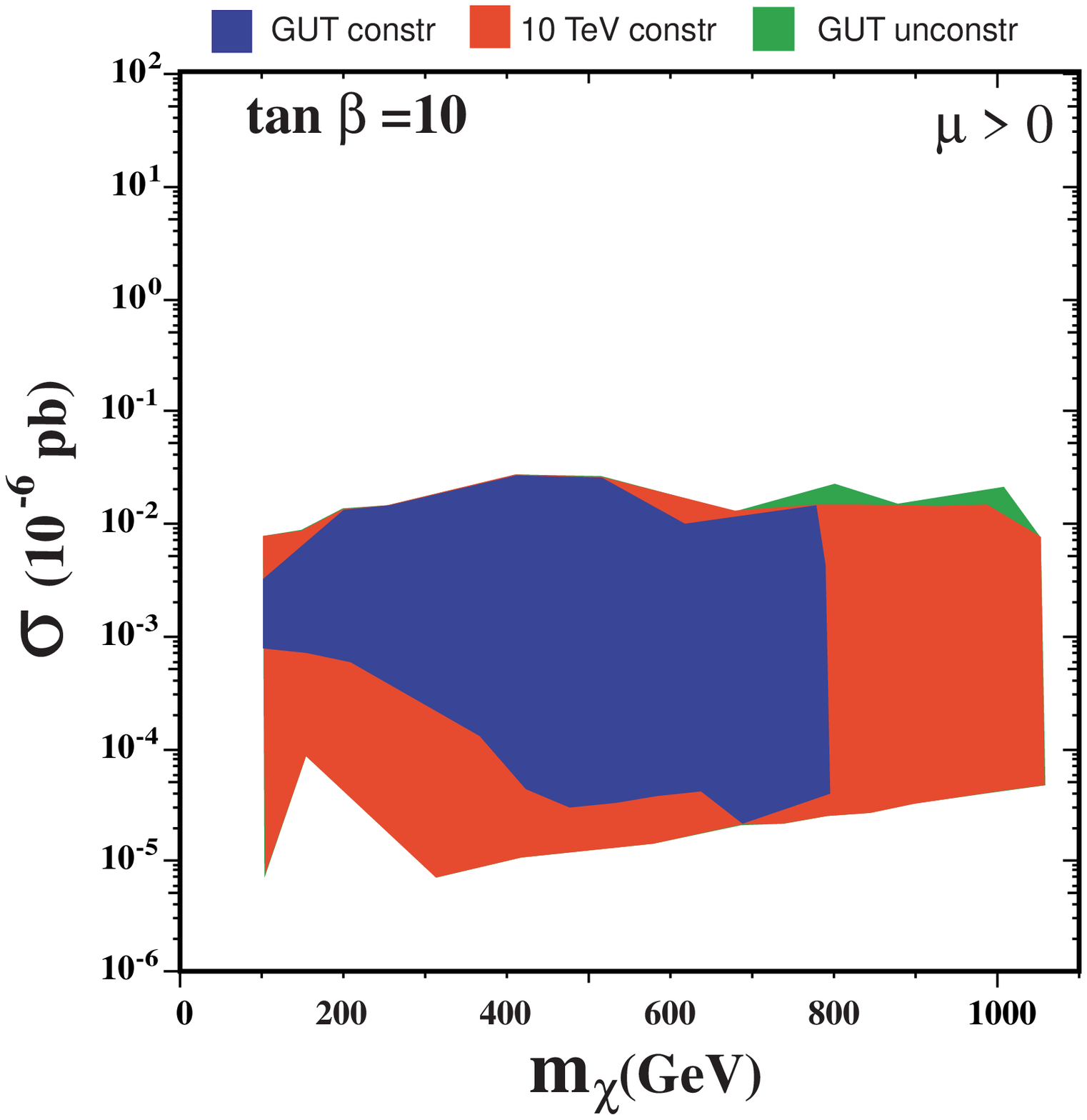,height=6.4cm}}
\vskip -0.1in
\end{center}
\begin{center}
\mbox{\epsfig{file=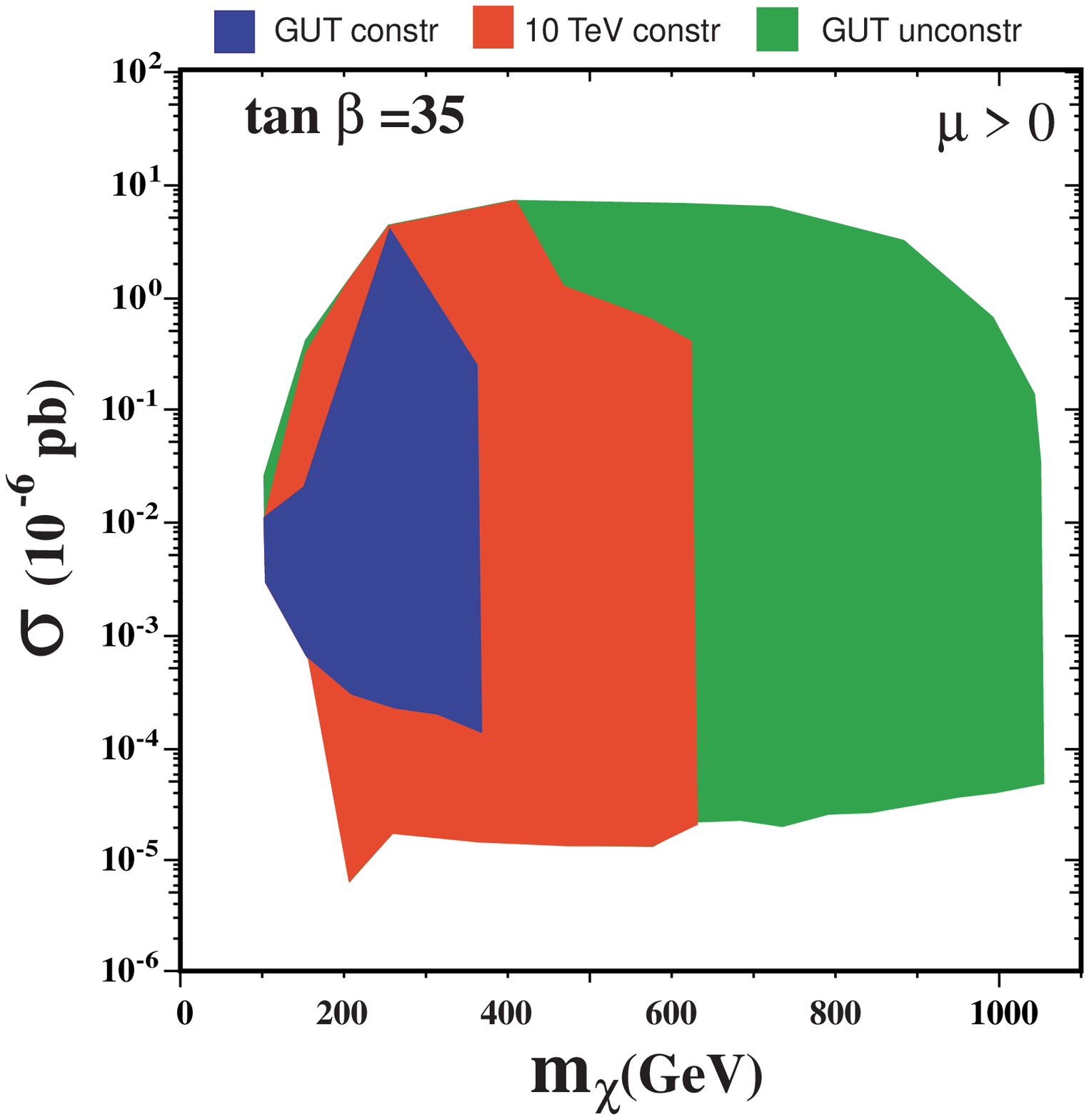,height=6.4cm}}
\hspace{0.4in}
\mbox{\epsfig{file=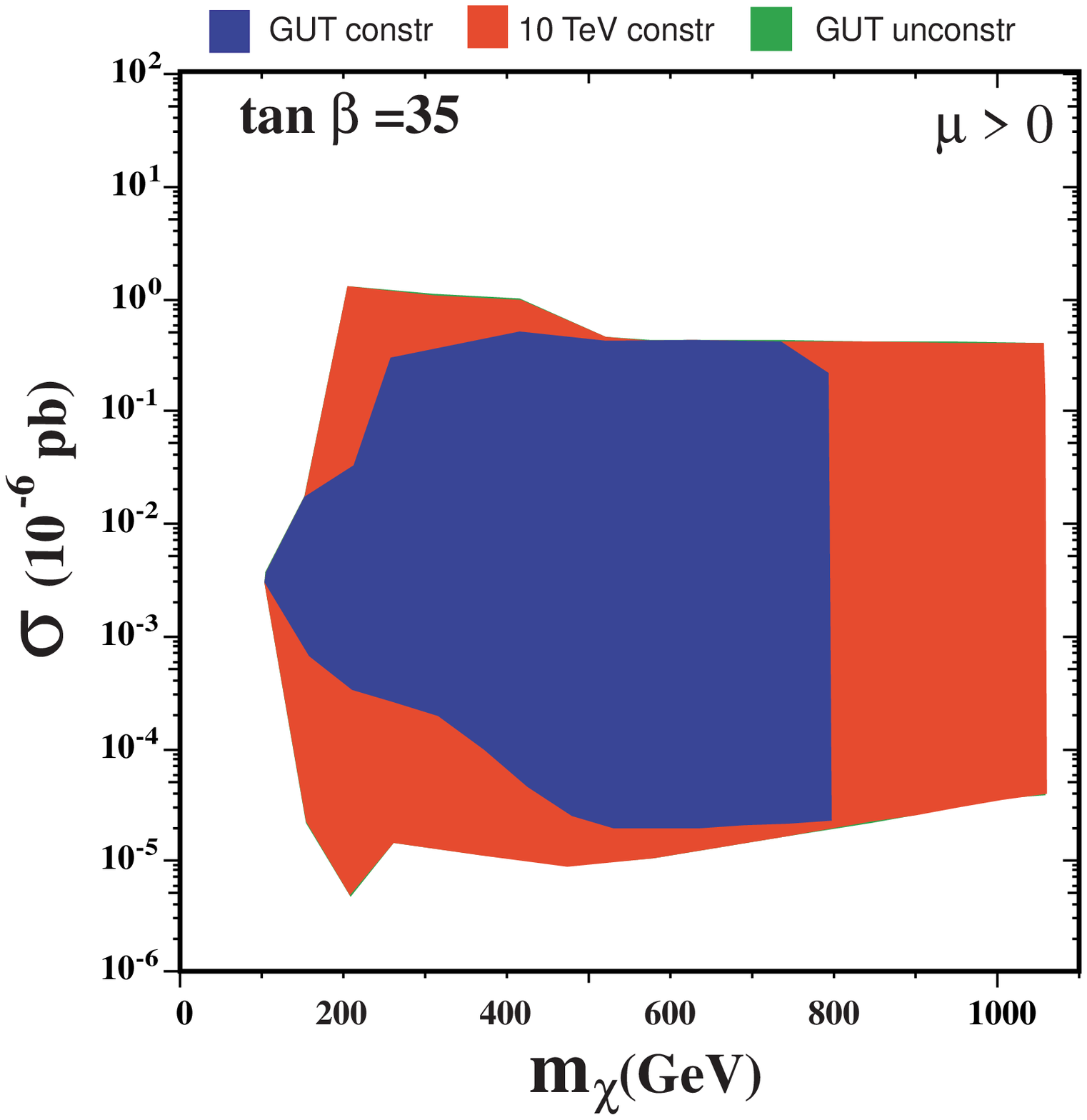,height=6.4cm}}
\vskip -0.1in
\end{center}
\begin{center}
\mbox{\epsfig{file=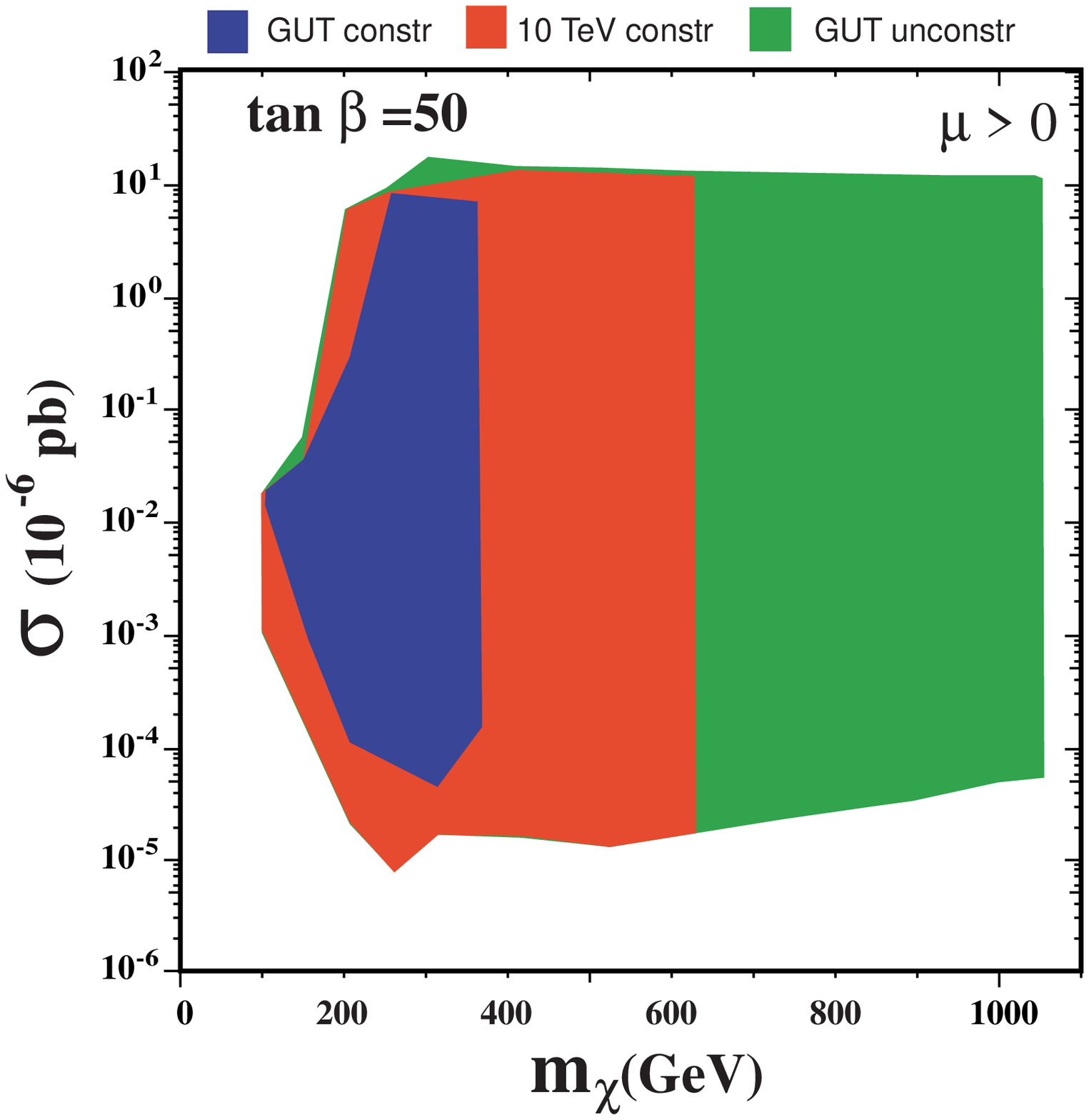,height=6.4cm}}
\hspace{0.4in}
\mbox{\epsfig{file=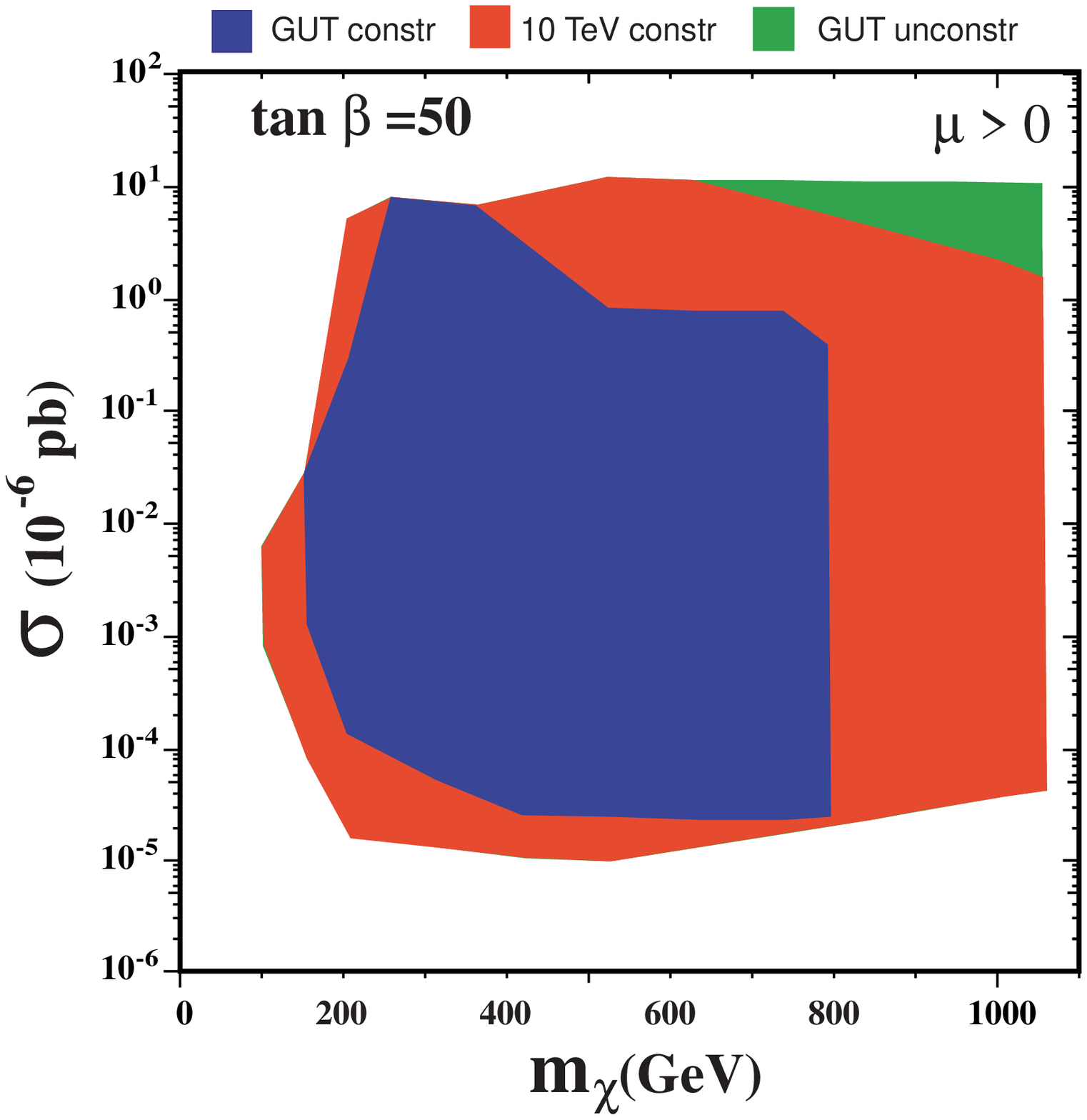,height=6.4cm}}
\end{center}
\vskip -0.2in
\caption{\label{fig:lowsc}\it
The spin-independent part of the $\chi-p$ cross section for $\tan \beta = 
10, 35$ and $50$ in the top, middle and bottom rows, respectively. The 
left (right) panels are for $200 \, \gev \leq
m_{\tilde q} \leq 2 \, \tev$ and $2 \, \tev \leq
m_{\tilde q} \leq 4 \, \tev$, respectively. The smallest (blue/dark) and 
medium (red)
shaded regions are obtained requiring no tachyons before the GUT and 
$10$~TeV scales, respectively. The largest (green/light) shaded regions 
are obtained without any such constraint, and correspond to a general 
LEEST analysis.}
\end{figure}

We discuss the spin-independent
neutralino-proton cross section first in the general LEEST
case, i.e., with parameters chosen in the low-energy effective MSSM and
ignoring the effects of RGE running. The results are shown as the most
extensive and lightest (green) shaded regions in Fig.~\ref{fig:lowsc}. There are
six panels, two each for $\tan \beta = 10, 35, 50$, in the top, middle and
bottom rows of panels in the figure. The left panels have $200 \, \gev
\leq m_{\tilde q} \leq 2 \, \tev$ and the right panels have $2 \, \tev
\leq m_{\tilde q} \leq 4 \, \tev$. We see that, as expected, the maximum
values of the cross section are higher in the left panels than in the
right panels. This indicates the importance of the $t$-channel
squark-exchange diagrams for the points with large cross sections, when
the exchanged squark masses are light. We also notice that larger values
of $\tan \beta$ have higher maximum cross sections, which is also known to
be true for the CMSSM case.  These results are in reasonably close 
agreement with those of~\cite{lowmssm}.

When we evolve the MSSM RGEs up and demand
that tachyonic sfermions be absent at the GUT scale $\sim 10^{16} \,
\gev$, we obtain the results shown as the smallest and darkest (blue)
shaded regions in Fig.~\ref{fig:lowsc}. In the left panels, these shaded
regions are cut at $\mchi \cong 400$ GeV, whilst in the right panels
they extend to
$\mchi \cong 800$ GeV. This is due to the fact that, as we evolve to
higher energies using the RGEs, the squark masses get smaller at rates
that depend on the gaugino masses, as described in section 3. 
The larger the gaugino masses, the
smaller the squark masses are at the GUT scale. As can be seen in
Fig.~\ref{fig:msql}, even if we start with $m_{\tilde q}(m_Z) = 2000$ GeV,
we find zero squark masses at the GUT scale for $m_{1/2}$ around 1000~GeV,
corresponding to a neutralino mass around 400 GeV. If we increase
the squark mass to 4000 GeV, $m_{1/2}$ can go up to around 2000~GeV,
corresponding to $m_\chi \sim 800$~GeV.

In the left panels of Fig.~\ref{fig:lowsc}, we observe that the maximum
cross section is generally lower than what would obtain if one did not
impose the GUT constraint. This reflects the fact that, for any given
value of $m_\chi$ and hence $m_{1/2}$, a lower limit on the low-energy
effective squark masses is imposed by requiring the absence of tachyonic
sfermions at the GUT scale. 
While this effect is quite pronounced at $\tan \beta = 10$, we note that the
reduction in the maximum cross section (for fixed $m_\chi$) is smaller
at higher $\tan \beta$.  
Furthermore, at low values of $m_\chi$ ($\sim$ 250 GeV), there is almost
no reduction at all. In the panels on the right, no reduction is
seen, as the squark masses are already so high that squark exchange is
no longer the dominant contribution to the cross section.

Even if one does not require the absence of tachyons up to the GUT scale,
their absence up to a lower effective scale still imposes important
constraints on the MSSM, as is also seen in Fig.~\ref{fig:lowsc}.  The
medium-sized and -shaded (red) regions show the ranges of the
spin-independent $\chi - p$
elastic cross section allowed if one forbids the existence of tachyons at
scales below 10~TeV rather than the GUT scale. We see that, even with this
much-reduced maximal scale, the RGEs provide significant restrictions.
In fact, most of the power of the
non-tachyonic constraint appears well below the GUT scale.

\begin{figure}
\begin{center}
\vskip -0.5in
\mbox{\epsfig{file=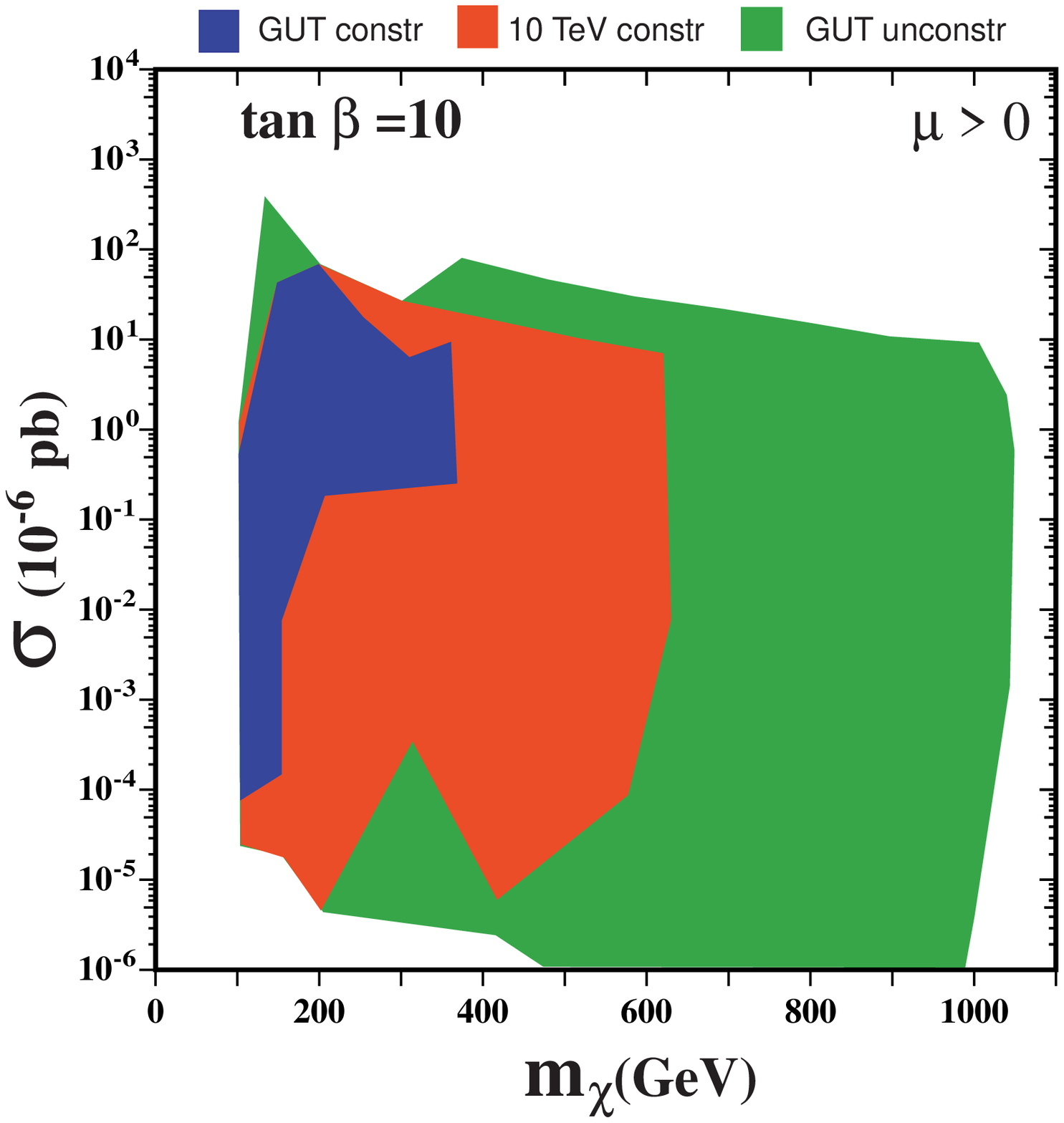,height=6.4cm}}
\hspace{0.4in}
\mbox{\epsfig{file=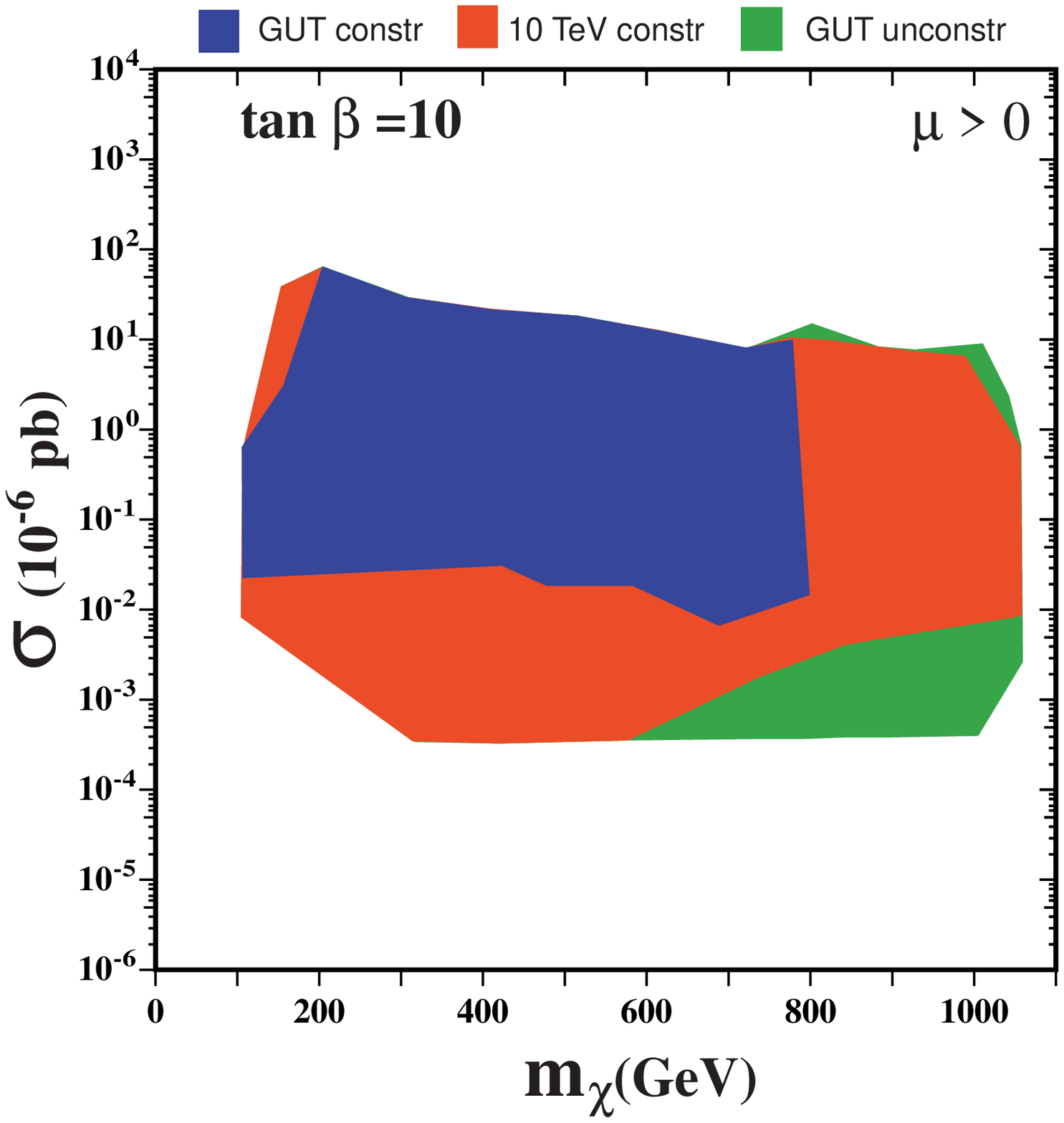,height=6.4cm}}
\vskip -0.1in
\end{center}
\begin{center}
\mbox{\epsfig{file=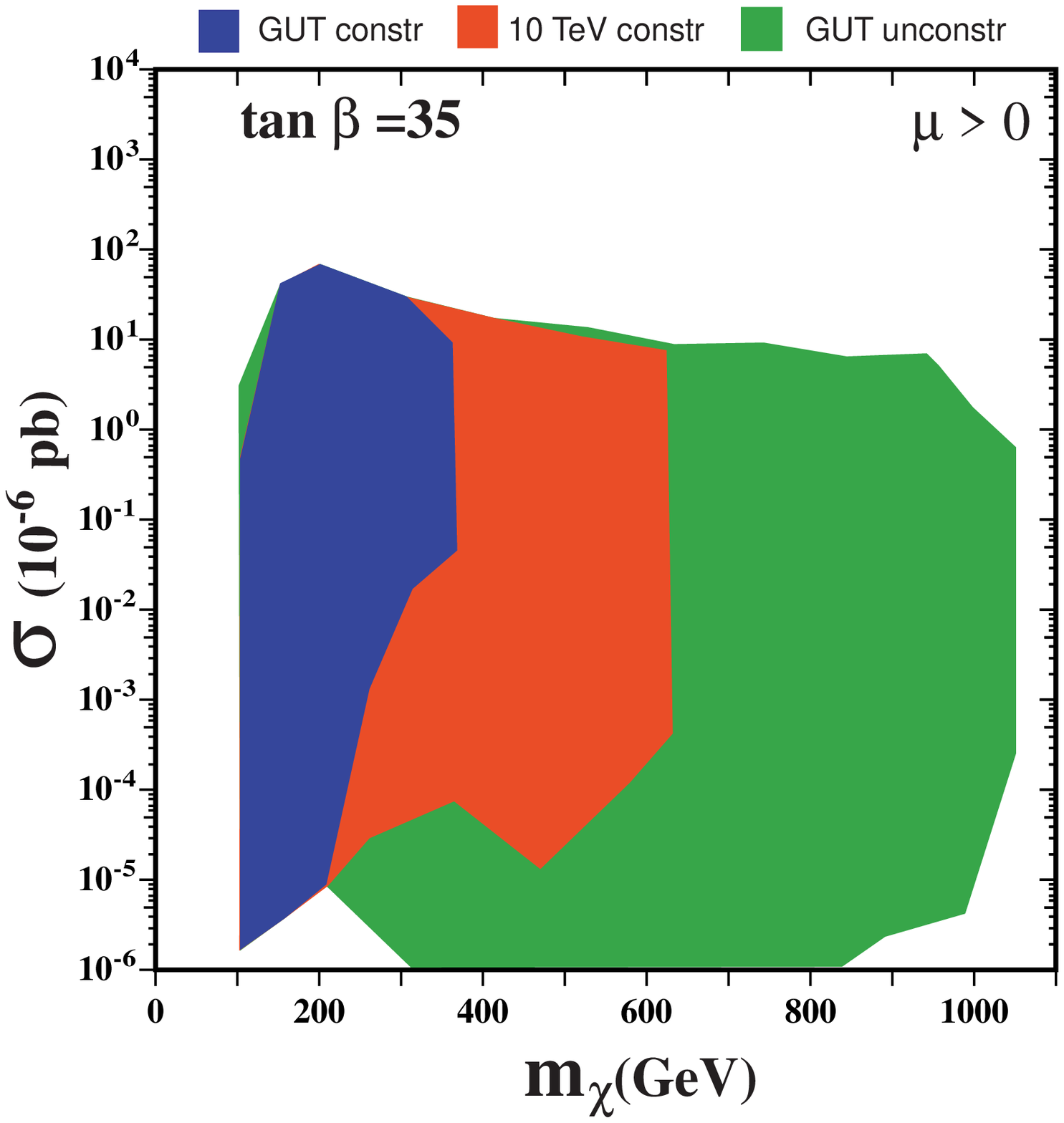,height=6.4cm}}
\hspace{0.4in}
\mbox{\epsfig{file=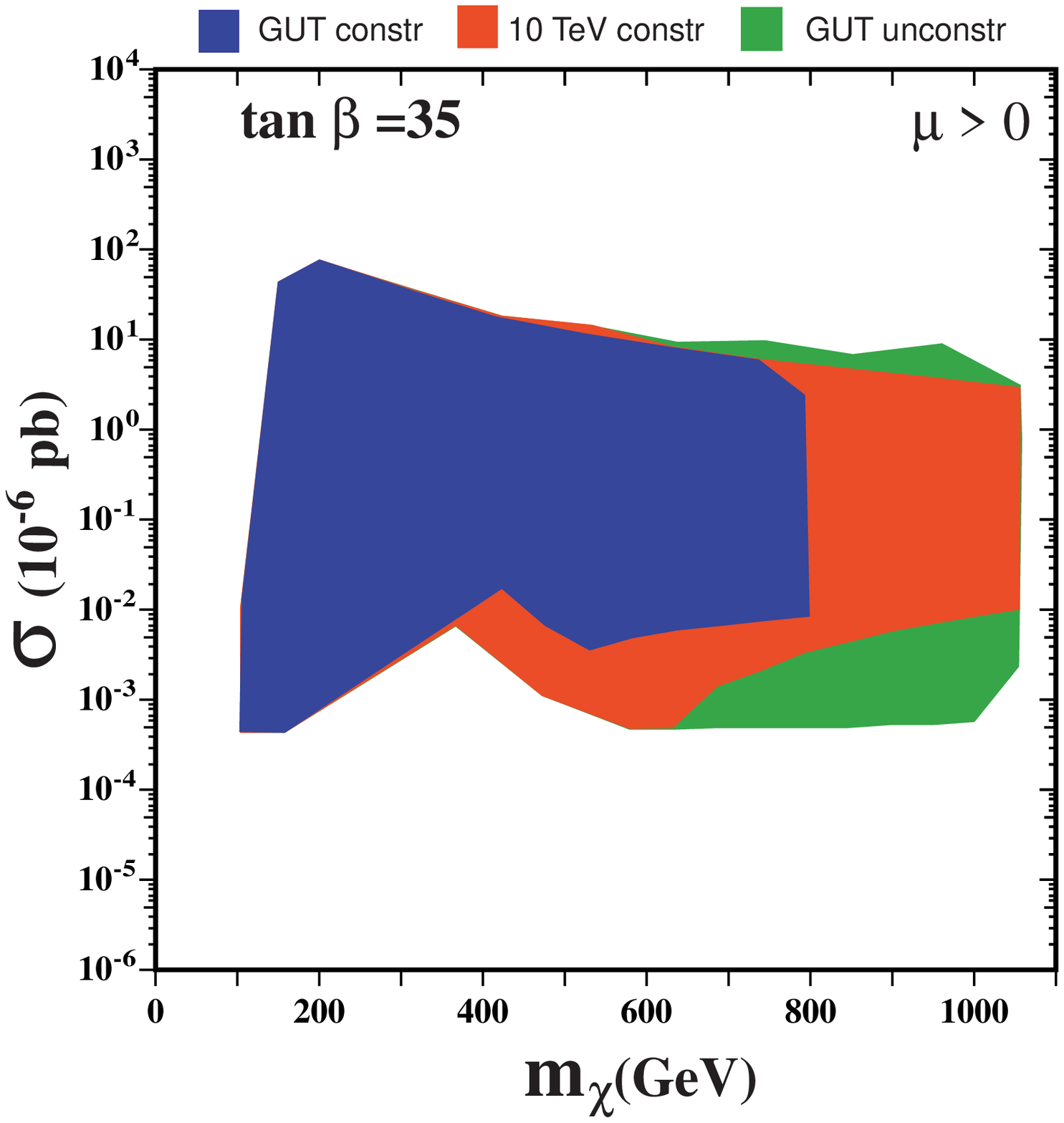,height=6.4cm}}
\vskip -0.1in
\end{center}
\begin{center}
\mbox{\epsfig{file=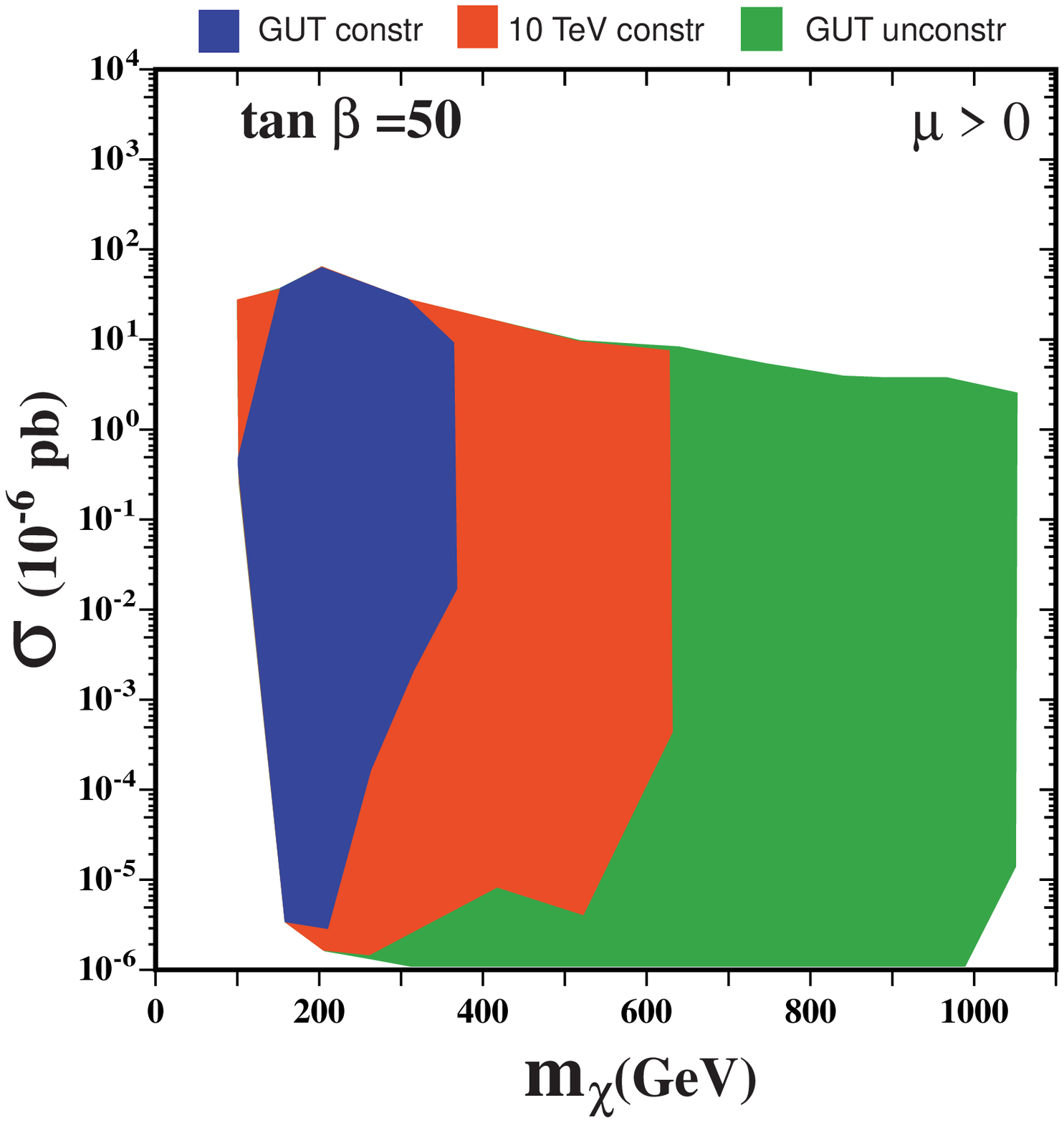,height=6.4cm}}
\hspace{0.4in}
\mbox{\epsfig{file=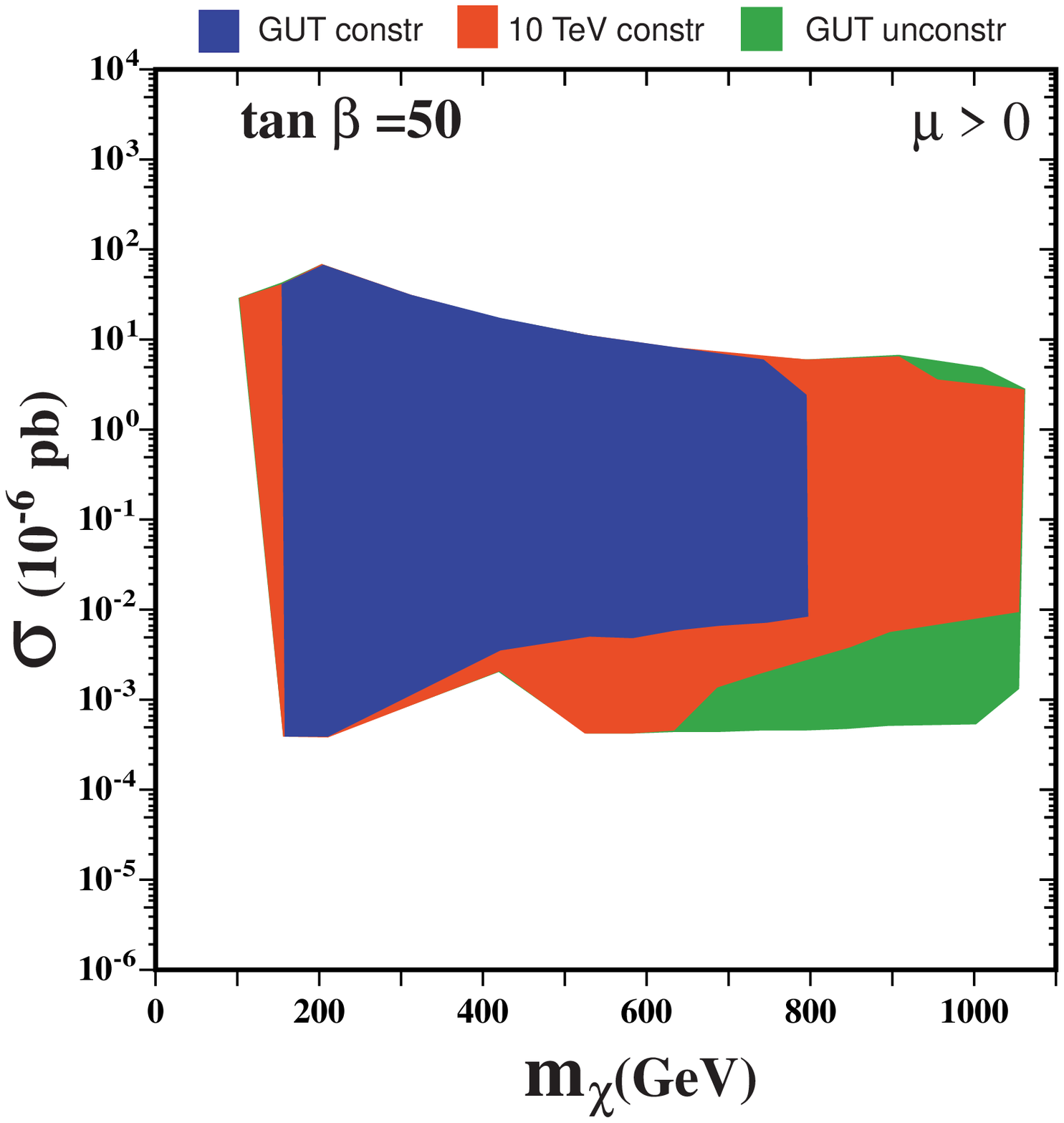,height=6.4cm}}
\end{center}
\vskip -0.2in
\caption{\label{fig:spinsc}\it
The spin-dependent part of the $\chi-p$ cross section for $\tan \beta = 
10, 35$ and $50$ in the top, middle and bottom rows, respectively. 
The various regions are as in Fig.~\ref{fig:lowsc}.}
\end{figure}

In Fig.~\ref{fig:spinsc} we show analogous results for the spin-dependent 
part of the $\chi - p$ elastic scattering cross section. Looking at the left or
the right panels only, we see that the maximum cross sections are not 
greatly
reduced by using the RGEs and requiring the absence of tachyons up to high 
energies.
However, if we combine the 
left panels with
the right ones, we nevertheless find that a smaller range of
cross sections is allowed when we apply the RGE
constraint than would be permitted without the RGE constraint, especially 
for large
$m_\chi$.

\begin{figure}
\begin{center}
\mbox{\epsfig{file=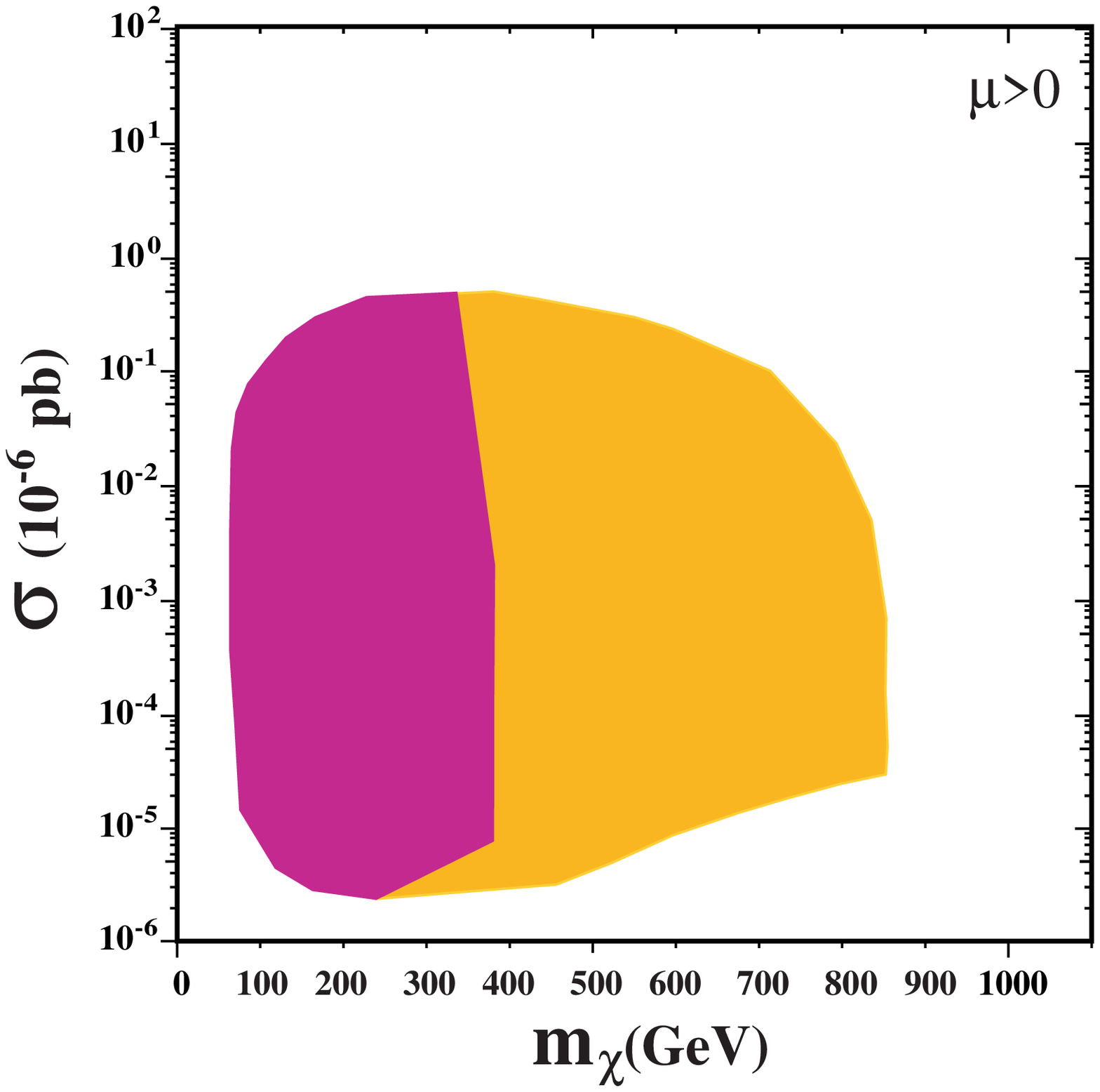,height=7cm}}
\hspace{0.2in}
\mbox{\epsfig{file=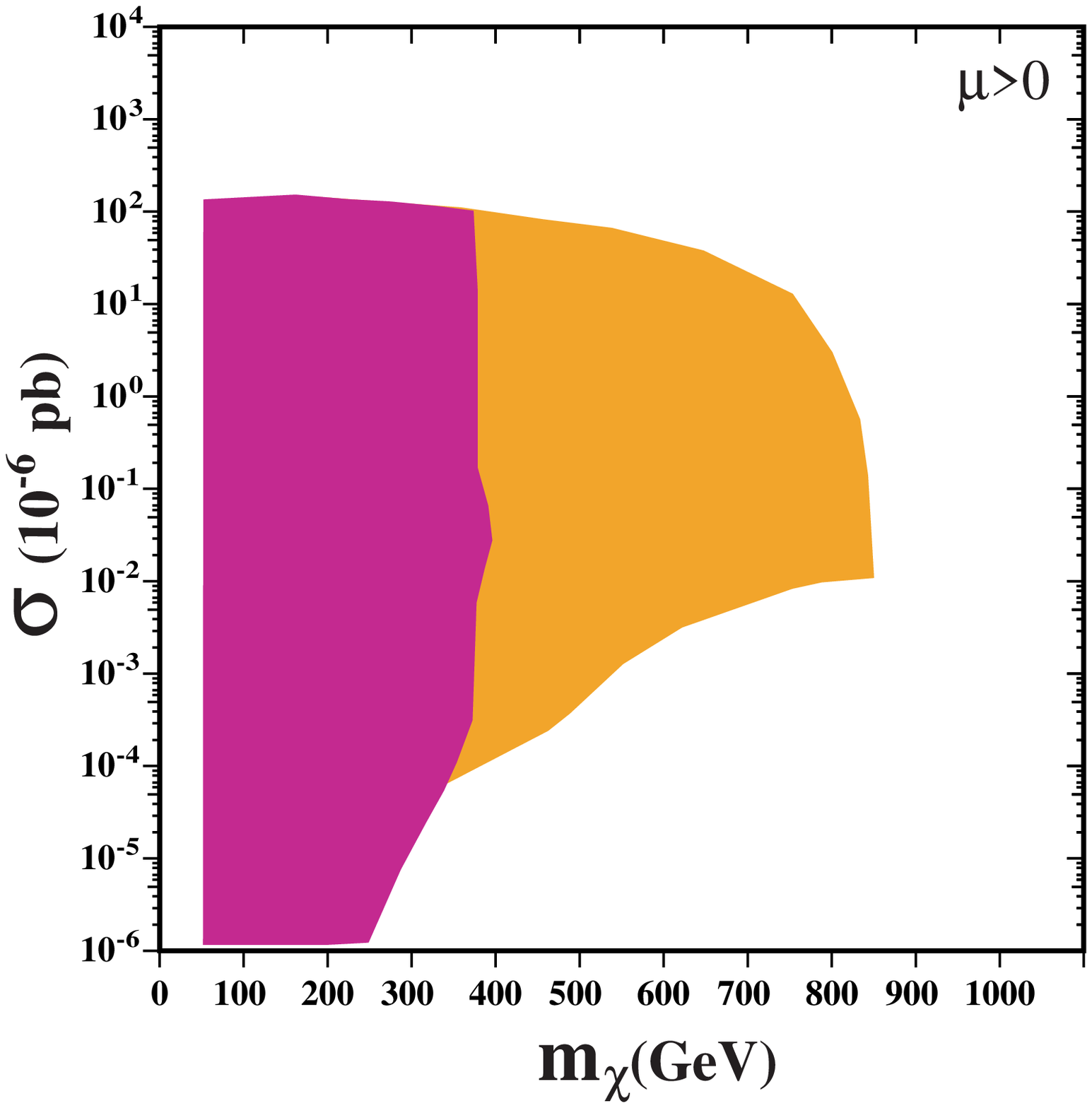,height=7cm}}
\end{center}
\caption{\label{fig:samh}\it
Left (right) panels: the ranges of the spin-independent (spin-dependent) 
$\chi-p$ cross sections found in a top-down approach, 
requiring that there is no tachyon at any scale from $M_{GUT}$ to $M_Z$ 
and combining all values of $\tan \beta$. 
The violet (dark) region 
corresponds to the low-mass range of squark masses 
$200 \gev < m_{\tilde q} < 2 \tev$, while the yellow (light) to the high
$2 \tev < m_{\tilde q} < 4 \tev$.
}
\end{figure}

Finally, we compare the above results with the range we find for the
$\chi-p$ cross section if we start from the GUT scale with the ranges of
input parameters described in Section~2, and run down from $M_{GUT}$ to $M_Z$.  
This top-down approach
gives us Fig.~\ref{fig:samh}, where all values of $\tan \beta$ are
combined. To facilitate the comparison with the previous figures,
we distinguish between the regions which give 
rise to squark masses above and below 2 TeV, corresponding to the right and left
panels of the previous two figures.  The area shown here
should be compared to the dark (blue) shaded region corresponding to the
GUT constraint. Indeed, the resulting ranges are similar to those found on
the left side (both sides) of Fig.~\ref{fig:lowsc}, when one imposes the
GUT constraint and combines all the three values of $\tan \beta$. 
Qualitatively, therefore, we find that the LEEST approach gives
similar results to the GUT (top-down) approach, so long as the
the tachyonic stability constraints are imposed.
We note,
however, that the allowed cross-section ranges do differ from those in the
CMSSM \cite{cmssmnp,npcross,othernp}, because the allowed parameter choices are no longer restricted to an
extreme part of the coannihilation tail and, even within this tail, the other weak-scale
sparticle masses have wider ranges than in the CMSSM.  On the other hand,  
these results are very similar to the NUHM, which is a relaxed version of the
CMSSM in which the Higgs soft masses are non-universal but the squark
and slepton masses are universal \cite{efloso}.  

\section{Conclusions}

We have compared in this paper predictions for neutralino-proton
scattering in models that are free from tachyonic instabilities all the
way up to the GUT scale, or just up to 10~TeV, with those of models in
which only a low-energy effective supersymmetric theory (LEEST) is
assumed, and no restriction is placed on its behaviour above the weak
scale. As we have shown, larger LSP masses and larger cross sections are
obtainable in the LEEST approach, but at the risk of sacrificing some of
the major motivations for postulating low-energy supersymmetry.
Specifically, we argue that a generic LEEST cannot be extrapolated up to
the GUT scale.  The main issue is whether the tachyonic instabilities that
are generic in the LEEST approach are tolerable.  Clearly in any theory
with GUT scale unification, they are not.  Cross sections larger than those in
the CMSSM may be attainable, particularly at large LSP
masses,  if such tachyonic instabilities are allowed.

\vskip 0.5in
\vbox{
\noindent{ {\bf Acknowledgments} } \\
\noindent 
We would like to thank M. Shifman and M. Srednicki for helpful conversations.
The work of K.A.O., Y.S. and V.C.S. was supported in part by DOE grant
DE--FG02--94ER--40823.}

\end{document}